\documentclass[12pt,preprint]{aastex}
\usepackage{graphicx}
\usepackage[]{natbib}

\def\dflux{$\cdot10^{-11}$~ph~cm$^{-2}$\,s$^{-1}$\,TeV$^{-1}$}
\def\arcsec{$''$}
\def\arcmin{$'$}
\shorttitle{Crab nebula and pulsar between 500~GeV and 80~TeV}
\shortauthors{Aharonian et al.}
\begin{document}
\title{The Crab Nebula and Pulsar between 500~GeV and 
80~TeV: Observations with the HEGRA stereoscopic air Cherenkov telescopes} 
\author{ F.~Aharonian\altaffilmark{1},
A.~Akhperjanian\altaffilmark{7},
M.~Beilicke\altaffilmark{4},
K.~Bernl\"ohr\altaffilmark{1,13},
H.-G.~B\"orst\altaffilmark{5},
H.~Bojahr\altaffilmark{6},
O.~Bolz\altaffilmark{1},
T.~Coarasa\altaffilmark{2}, 
J.L.~Contreras\altaffilmark{3}, 
J.~Cortina\altaffilmark{2,10}, 
S.~Denninghoff\altaffilmark{2}, 
M.V.~Fonseca\altaffilmark{3}, 
M.~Girma\altaffilmark{1}, 
N.~G\"otting\altaffilmark{4}, 
G.~Heinzelmann\altaffilmark{4}, 
G.~Hermann\altaffilmark{1}, 
A.~Heusler\altaffilmark{1}, 
W.~Hofmann\altaffilmark{1}, 
D.~Horns\altaffilmark{1,14},
I.~Jung\altaffilmark{1}, 
R.~Kankanyan\altaffilmark{1}, 
M.~Kestel\altaffilmark{2}, 
A.~Kohnle\altaffilmark{1}, 
A.~Konopelko\altaffilmark{1,13}, 
D.~Kranich\altaffilmark{2}, 
H.~Lampeitl\altaffilmark{4}, 
M.~Lopez\altaffilmark{3}, 
E.~Lorenz\altaffilmark{2}, 
F.~Lucarelli\altaffilmark{3}, 
O.~Mang\altaffilmark{5}, 
D.~Mazin\altaffilmark{2}, 
H.~Meyer\altaffilmark{6}, 
R.~Mirzoyan\altaffilmark{2}, 
A.~Moralejo\altaffilmark{3}, 
E.~O\~{n}a-Wilhelmi\altaffilmark{3}, 
M.~Panter\altaffilmark{1}, 
A.~Plyasheshnikov\altaffilmark{1,8},
G.~P\"uhlhofer\altaffilmark{1,11},
R.~de\,los\,Reyes\altaffilmark{3}, 
W.~Rhode\altaffilmark{6}, 
J.~Ripken\altaffilmark{4}, 
G.~Rowell\altaffilmark{1}, 
V.~Sahakian\altaffilmark{7}, 
M.~Samorski\altaffilmark{5}, 
M.~Schilling\altaffilmark{5}, 
M.~Siems\altaffilmark{5}, 
D.~Sobzynska\altaffilmark{2,9},
W.~Stamm\altaffilmark{5}, 
M.~Tluczykont\altaffilmark{4,12},
V.~Vitale\altaffilmark{2}, 
H.J.~V\"olk\altaffilmark{1}, 
C.~A.~Wiedner\altaffilmark{1},
W.~Wittek\altaffilmark{2}}
\altaffiltext{1}{Max-Planck-Institut f\"ur Kernphysik,
Postfach 103980, D-69029 Heidelberg, Germany}
 
\altaffiltext{2}{Max-Planck-Institut f\"ur Physik, F\"ohringer Ring
6, D-80805 M\"unchen, Germany}
 
\altaffiltext{3}{
Universidad Complutense, Facultad de Ciencias
F\'{\i}sicas, Ciudad Universitaria, E-28040 Madrid, Spain }
 
\altaffiltext{4}{Universit\"at Hamburg, Institut f\"ur
Experimentalphysik, Luruper Chaussee 149,
D-22761 Hamburg, Germany}
 
\altaffiltext{5}{
Universit\"at Kiel, Institut f\"ur Experimentelle und
Angewandte Physik,
Leibnizstra{\ss}e 15-19, D-24118 Kiel, Germany}
 
\altaffiltext{6}{
Universit\"at Wuppertal, Fachbereich Physik,
Gau{\ss}str.20, D-42097 Wuppertal, Germany}
 
\altaffiltext{7}{
Yerevan Physics Institute, Alikhanian Br. 2, 375036
Yerevan, Armenia}
 
\altaffiltext{8}{
On leave from  
Altai State University, Dimitrov Street 66, 656099 Barnaul, Russia}
 
\altaffiltext{9}{
Home institute: University Lodz, Poland}
 
\altaffiltext{10}{
Now at Institut de F\'{\i}sica d'Altes Energies, UAB, Edifici Cn, E-08193, Bellaterra (Barcelona), Spain}
 
\altaffiltext{11}{
Now at Landessternwarte Heidelberg, K\"onigstuhl, Heidelberg, Germany}
 
\altaffiltext{12}{
Now at Laboratoire Leprince-Ringuet, Ecole Polytechnique, Palaiseau, France (IN2P3/CNRS)}

\altaffiltext{13}{
 Now at Humboldt Universit\"at, Institut f. Physik, Newtonstr.~15, Berlin, Germany}
 
\altaffiltext{14}{
Corresponding author: D.~Horns \email{Dieter.Horns@mpi-hd.mpg.de}}
\tighten

%%\date{Received / Accepted}
%%\offprints{\\D.Horns, \email{Dieter.Horns@mpi-hd.mpg.de}}

\begin{abstract}
 The Crab supernova remnant has been observed regularly with the stereoscopic system of 5
 imaging air Cherenkov telescopes that was part of the High Energy Gamma Ray
 Astronomy (HEGRA) experiment. In total, close to 400 hours of useful data have
 been collected from 1997
 until 2002. The differential energy spectrum of the
 combined data-set can be approximated by a power-law type energy spectrum:
 $d\Phi/dE=\Phi_0\cdot (E/\mathrm{TeV})^{\Gamma}$,
 $\Phi_0=(2.83\pm0.04_\mathrm{stat}\pm0.6_\mathrm{sys})$~\dflux\ and
 $\Gamma=-2.62\pm0.02_\mathrm{stat.}\pm{0.05}_\mathrm{sys.}$. The spectrum
 extends up to energies of 80~TeV and is well matched by model calculations
 in the framework of inverse Compton scattering of various seed photons in the
 nebula including for the first time a recently detected 
compact emission region at mm-wavelengths. 
The {observed indications for a} gradual steepening of the energy spectrum in data is
expected in  the inverse Compton emission model.  
The average magnetic field in the emitting volume is determined
 to be $(161.6\pm0.8_\mathrm{stat}\pm18_\mathrm{sys})~\mu$G.
 The presence of protons in the nebula is not required to explain the observed 
 flux and upper limits on the injected power of protons are calculated 
 being as low as 20~\% of the total spin down luminosity for bulk 
 Lorentz factors of the wind in the range of $10^4-10^6$.
 The position and size of the emission region have been studied over a
 wide range of energies. The position is 
 shifted by 13\arcsec\ to the west of the pulsar with a systematic
uncertainty of 25\arcsec. No significant shift in the position with energy  is
observed. The size of the
 emission region is constrained to be less than 2\arcmin\ at
 energies between 1 and 10~TeV. Above 30~TeV
 the size is constrained to be less than 3\arcmin.  
 No indications for pulsed emission has been found and
 upper limits in differential bins of energy have been calculated reaching
 typically 1-3~\% of the unpulsed component.  
\end{abstract}

 \section{Introduction}
  Observations of the Crab pulsar and nebula have been carried out in every
accessible  wavelength band (ground and space based). The source has been
established as a TeV emitter with the advent of ground based Cherenkov
imaging telescopes with sufficient sensitivity \citep{1989ApJ...342..379W}. 
The
broad band spectral energy distribution (SED) is exceptionally
complete in coverage and unique amongst all astronomical objects observed and studied. The
historical lightcurve of the supernova explosion of 1054 A.D. is not
conclusive with respect to the type of progenitor star  and 
leaves many questions concerning the stellar evolution of the progenitor star unanswered \citep{stephenson}. At the
present date, the observed remnant is of a plerionic type with a bright
continuum emission and filamentary structures emitting mainly in lines
in the near infrared and optical range.  

% Power available
 The remaining compact central object is a pulsar with a period of 33~ms and a
spin-down luminosity of ($\propto \dot P^3/P)~
 5\cdot 10^{38}$~erg~s$^{-1}$.  The emitted
power of the continuum peaks in the hard UV/ soft X-ray at $\approx
10^{37}$~erg~s$^{-1}$ (assuming a distance of 2~kpc).  Given the kinetic energy of the
spinning pulsar as the only available source of energy in the system, the
spin-down luminosity is efficiently converted into radiation. There is no
evidence for accretion onto the compact object as an alternative mechanism to
feed energy into the system \citep{2004ApJ...601L..71B}. 

% SED
  The emission of the Crab nebula is predominantly produced by nonthermal
processes (mainly synchrotron and inverse Compton) 
which form a continuum emission covering the range from radio to
very-high energy gamma-rays. The synchrotron origin of the optical and radio continuum emission was 
proposed by \citet{shklovskii} 
and experimentally identified by polarization measurements \citep{dombrovsky}.

 The general features of the observed SED are characterised
by 3 breaks connecting 4 power-law type spectra. The breaks occur in the near
infrared, near UV, and hard  X-ray. The break frequencies are inferred indirectly 
because they occur in spectral bands which are difficult to observe.   The high 
energy part of the synchrotron spectrum cuts off at a few MeV. This
energy is very close to the theoretical limit for  synchrotron emission derived from
classical electrodynamics to 
be $h\nu_{max}= \kappa m_e\,c^2/\alpha\approx\kappa\cdot 68$~MeV (independent of the magnetic field)
with $\kappa\approx 1\ldots3$ depending on the broad-band spectrum of the emitting electrons \citep{2000NewA....5..377A}.  
In this sense, the Crab accelerates particles close to the possible limit.
The synchrotron origin of multi-MeV photons
requires the presence of electrons up to PeV energies. 
These PeV electrons inevitably radiate via inverse Compton scattering at photon 
energies up to and beyond 50~TeV with detectable fluxes
\citep{1998ApJ...492L..33T,ICRC2003}.
 Another independent indication for a common origin of
MeV and 50-100~TeV photons is possibly correlated variability of the
nebula emission in these two energy bands which has been discussed by \citet{horns}.

 In addition to the 
inverse Compton emission, other additional
mechanisms might contribute to the VHE luminosity:
Photons from $\pi^0$ decay from ions responsible for the acceleration
of positrons at the termination shock \citep{1995mpds.conf..257A} 
and bulk Comptonization of the relativistic wind \citep{2000MNRAS.313..504B}.

% Particle acceleration
 The origin of the emitting particles is commonly assumed to be the pulsar
which efficiently accelerates a relativistic wind of particles
(electrons/positrons and ions) in the proximity of the magnetosphere  with a
luminosity close to its spin-down luminosity.  The relativistic wind terminates
in a standing reverse shock which is commonly associated with a ring like
feature identified at a distance of $\approx 4 \cdot 10^{17}$~cm or 14$''$ by
recent X-ray observations \citep{2000ApJ...536L..81W}. At the position of the
termination shock, the flow is dominated by kinetic energy. 
The orientation of the toroidal magnetic field is transverse to the flow
direction and parallel to the shock surface \citep{1974MNRAS.167....1R}. 
This shock geometry is not in favor of efficient acceleration of particles
in shock drift or diffusive shock acceleration.  Two possible alternative 
acceleration mechanisms have been discussed in detail in the literature:

 A possible
mechanism to accelerate particles in the relativistic wind is the dissipation
of magnetic energy via reconnection of current sheets in the striped pulsar
wind \citep{1990ApJ...349..538C}, which was recently revisited by
\citet{2003ApJ...591..366K}.  This mechanism was found to be close to the
requirement to convert sufficiently fast  the initially Poynting flux dominated
flow. However, it requires a rather high pair injection rate of
$3\cdot10^{40}$~s$^{-1}$ and subsequently a small bulk Lorentz factor of the
flow of $<10^{4}$. 

An alternative process to 
efficiently accelerate particles at the shock suggested by \citet{1995mpds.conf..257A}
 invokes resonant scattering of
positrons by cyclotron waves induced by the ions in the downstream
region. Numerical particle-in-cell simulations \citep{1992ApJ...390..454H} 
show that this mechanism produces a power-law type spectrum. The ions are expected to produce gamma-rays
after interacting with the matter in the nebula \citep{2003A&A...405..689B,2003A&A...402..827A} 
and produce signatures in the observed gamma-ray spectrum.

% MHD
Independent of the actual acceleration mechanism, the injected nonthermal particle distribution in the
nebula 
expands and cools adiabatically 
\citep{1973ApJ...186..249P} in a subsonic flow until it
terminates at the envelope of the nebula.
 In the framework of the magneto-hydrodynamic (MHD) flow model of
\citet{1984ApJ...283..694K}, 
basic constraints on the bulk Lorentz factor of the wind and the
pulsar's luminosity have been derived.  It is noteworthy that in
these models the radio emission is not explained and an ad hoc assumption
about an extra population of radio emitting particles is required
\citep{1984ApJ...283..694K,1996MNRAS.278..525A}. 
 
Furthermore, assuming that the MHD approximation holds,
the downstream particle distribution, flow velocity, and magnetic field
strength can be derived.  Overall, the predicted synchrotron emission and its
angular dimension in this model are in good agreement with observations
\citep{1984ApJ...283..710K,1996MNRAS.278..525A}. 
 
% Model calculations for the inverse Compton emission of the nebula are presented in 
%Section~\ref{section:model} and compared to the energy spectrum (see Section~\ref{subsec:Energy_spectrum})
%in Section~\ref{section:Comparison}.

%Morphology
 The  morphology of the nebula is resolved at most wavelengths. It is 
characterized by a complex structure and
shows a general decrease of the size of the emission region with increasing
energy.  This is consistent with the decrease of the lifetime of electrons injected
into the nebula with increasing energies.  However, there are indications that at mm-wavelengths, a compact,
possibly non-thermal emission region is present that does not follow the
general behaviour \citep{2002A&A...386.1044B}. 
The central region close to the torus-like structure is also
known to exhibit temporal variability at  Radio \citep{2001ApJ...560..254B}, 
optical \citep{1995ApJ...448..240H}, and  X-ray frequencies \citep{2002ApJ...577L..49H}. The
morphology of these moving structures first seen in optical observations 
has lead to the commonly used term
``wisps'' \citep{1969ApJ...156..401S} which are possibly (for a discussion
see also Hester et al., 1995) explained by ions penetrating downstream
 of the standing shock and accelerating positrons and electrons
via cyclotron resonance \citep{1994ApJ...435..230G}.  
Whereas at radio wavelengths the measured spectrum is spatially
constant over the  entire nebula, at higher frequencies, a general softening of the
spectrum from the central region to the outer edge of the nebula is observed. This 
tendency is also explainable with the short life time of the more energetic electrons.
Recent high resolution spectral imaging at X-ray energies have revealed a
\textit{hardening} of the spectrum in the torus region well beyond the
bright ring-like structure that is usually identified with the standing relativistic
shock \citep{2004astro.ph..3287M}. The spectral hardening at the torus could be
indicative for ongoing acceleration far from the pulsar.
 In this case, the
predictions for the size of the TeV emission region based upon injection of electrons at the
supposed standing reverse shock wave at 14\arcsec\ projected distance are very likely underestimating
the true size. The torus is located at $\approx 45$\arcsec\ projected distance to the pulsar. 
Additional components as e.g. photons from $\pi^0$ decay are expected to have roughly the same
angular size as the radio nebula ($\approx 3$\arcmin) which is resolvable with stereoscopic ground based
Cherenkov telescopes.
% The results of analysing the morphology of the TeV emission region are
%presented in Section~\ref{section:morphology_results}.

%Pulsar
 Besides the nebula discussed above, the \textit{pulsar} is a prominent source of pulsed radiation  up to GeV energies
\citep{1977A&A....61..279B,1993ApJ...409..697N}.
The characteristic bimodal pulse shape is retained over all wavelengths with variations of the relative 
brightness of the two main components. The pulsar has not been detected at energies beyond
10~GeV. Polar cap \citep{1978ApJ...225..226H} and outer gap \citep{1986ApJ...300..500C}  models have been successful at explaining the general
features of the pulsed emission as curvature radiation of an energetic electron/positron
plasma. The emission at TeV energies strongly depends on the location of the emission region. 
Generally, within polar cap models the optical depth for TeV photons is very large because of magnetic pair production
processes. Effectively, the polar cap region is completely opaque to TeV emission. 
Within outer gap models, inverse Compton scattered photons could be visible even at 
TeV energies. Again, the optical depth depends on the position of the outer gap in the magnetosphere. 
In any case, strong variation of the optical depth with energy results in rather narrow features in energy
for a possible signal in a $\nu f_\nu$ diagram. This motivates a dedicated analysis searching
for signals in narrow differential energy bins. 

The paper is structured in the following way: After a description of the instrument (Section~\ref{section:experiment})
and data analyses (Section~\ref{section:data}),
 results on the energy spectrum, 
%(Section~\ref{subsec:Energy_spectrum}) 
morphology, 
%(Section~\ref{section:morphology_results}),
and pulsed emission (Section~\ref{section:phase_result}) are given in Section 4. Section~\ref{section:model} describes model calculations
of the inverse Compton emission and compares the prediction with the measured energy spectrum.% (Section~\ref{section:Comparison}). 
The average magnetic field in the
emission region is derived, 
%(Section~\ref{section:average_magnetic}), 
upper limits on the presence of ions in the wind are 
calculated, 
%(Section~\ref{section:ions}), 
and additional components are discussed.
 Finally, a summary and conclusion are given (Section~\ref{section:summary}).

\section{Experimental set-up}
\label{section:experiment}
 Observations were carried out with the HEGRA (High Energy Gamma Ray Astronomy)
stereoscopic system of five imaging air Cherenkov  telescopes located on the Roque de los
Muchachos on the Canary island of La Palma at an altitude of 2,200~m above sea
level.  The system consisted of optical telescopes each equipped with a
8.5~m$^2$ tessellated mirror surface arranged in a Davies-Cotton design.   The
horizontal (altitude/azimuth) mounts of the telescopes were computer controlled and could track objects in celestial
coordinates with an accuracy of better than 25\arcsec\
\citep{2003APh...20.267}.  The telescopes were arranged on the corners of a square
with a side length of 100~m with a fifth telescope positioned in the center.  
The individual mirror dishes collected Cherenkov light generated by charged particles
from air showers in the atmosphere which produces an image of the short (a few
nsec duration) light pulse in the focal plane. The primary focus was equipped
with an ultra-fast array of 271 photo-multiplier
tubes (PMT) arranged in a hexagonal shape at 0.25$^\circ$ spacing. The light
collection area of the PMTs was enhanced by funnels installed in front of the
photo-cathode.  The camera covered a field of view with a full opening angle of
4.1$^\circ$ in the sky suitable for surveys
\citep{2002JPhysG...28..2755,2003A&A...410..389R,ICRC2003...OG2.2.2319}. 
The feasibility of survey type observations is demonstrated by the 
first serendipitous discovery of an unidentified object with this instrument
named TeV~J2035+4130 \citep{2002A&A...393L..37A}.
 
% Stereoscopy
 The imaging air Cherenkov technique initially explored with single telescopes
is improved by adding stereoscopic information to the event reconstruction.
This was pioneered by the HEGRA group (see \citet{1993gamma...81,1996APh...5...119} and
below). The stereoscopic observation improves 
the sensitivity by reducing background  and achieving a low energy threshold with  a
comparably small mirror surface by efficiently rejecting local muon background
events. An important benefit of the stereoscopy is also the redundancy of
multiple images allowing to control efficiently systematic uncertainties.
Additionally, the reconstruction of the energy and the 
direction of the shower is substantially improved with respect to single-telescope
imaging \citep{1999APh....12..135H,2000APh....12..207H}.  
Moreover, the probability of night sky background triggers in two
telescopes in coincidence is strongly reduced.  The trigger condition for
the telescope requires two adjacent pixels exceeding 6 photo electrons (p.e.)
within an effective gate time of 14~nsec.

 Further details on the telescope hardware are given in \citet{padova}, a
 description of the trigger and read-out electronics can be found in
 \citet{1998APh...8...223}.

 The telescope system with 4 (from 1998 onwards 5) telescopes was
 operational between spring 1997 and fall 2002. 

%
%
%
%
%

%%%%%%%%%%%%%%%%%%%%%%%%%%%%%%%%%%%%%%%%%%%%%%%%%%%
%%%%%%%%%%%%%%%%%%%%%%%%%%%%%%%%%%%%%%%%%%%%%%%%%%%
\section{Data analyses}
%%%%%%%%%%%%%%%%%%%%%%%%%%%%%%%%%%%%%%%%%%%%%%%%%%%
%%%%%%%%%%%%%%%%%%%%%%%%%%%%%%%%%%%%%%%%%%%%%%%%%%%
\label{section:data}

\subsection{Data taking}
  Observations are taken  in a ``wobble'' observation mode, in  which the
telescopes point at a position shifted by 0.5$^\circ$ in declination to the
position of the Crab pulsar. The sign of the shift is alternated every 20~min.
All telescopes are aligned at the same celestial position and track the object
during its transit in the sky.  The wobble observation mode efficiently
increases the available observation time while decreasing systematic
differences seen between dedicated ON- and OFF-source observations. The
background for point-like sources is estimated by using 5 OFF-source positions
with a similar acceptance as for the ON-source observations
\citep{2001A&A....370..112}.

  In the observational season 1998/1999, dedicated observations at larger
zenith angles (z.a. distance $>45^\circ$) were performed in order to increase
the collection area for high energy events (typically above 10~TeV). All
observations have been carried out under favorable conditions with clear skies,
low wind speeds ($<10$~m~$s^{-1}$), and  a relative humidity below 100~\%. 

The overall relative operation efficiency of the HEGRA system of Cherenkov
telescopes reached 85~\% {of the total available darkness time} 
such that 5,500 hours of data on more than 100
objects and various scans \citep{2001A&A...375.1008A,2002A&A...395..803A} were
collected in 67 months of operation.
% The available \textit{darkness} time
%(moonless time with astronomical darkness condition i.e. sun 18$^\circ$ below
%horizon) for the entire operational period amounts to $\approx 8800$~hrs. The
%required conditions for a safe and useful operation (relative humidity below
%100~\%, average wind-speeds below 10~m~s$^{-1}$) of the telescope reduces the
%\textit{usable} time to $\approx 6500$~hrs of which approximately $5500$~hrs
%were actually observed.  
La Palma can be considered an excellent  site for
Cherenkov light observations and moreover the HEGRA system of Cherenkov
telescopes was functioning very smoothly and reliably.  The entire data-set
collected by the HEGRA system of telescopes   has been used  to search for
serendipitous sources in the field of view \citep{ICRC2003...OG2.2.2319}.
Dedicated test observations of the Crab nebula with a topological trigger,
reduced energy threshold, and convergent pointing mode are reported elsewhere
\citep{2003APh...19..339}.  

%%%%%%%%%%%%%%%%%%%%%%%%%%%%%%%%%%%%%%%%%%%%%%%%%%%
\subsection{Data selection}
%%%%%%%%%%%%%%%%%%%%%%%%%%%%%%%%%%%%%%%%%%%%%%%%%%%

% Telescope set-ups
Throughout the time of operation, the data acquisition
\citep{ICRC1997....OG10.3.20}  and electronics setup of the readout of the
cameras and central triggering  have remained unchanged. As a consequence of
the upgrade from 4 to 5 telescopes in 1998 and occasional technical problems,
the number of operational telescopes varies from 3 (5.5~\% of the total
operational time) to 4 (38~\%) and 5 instruments (56~\%). Runs with less than 3
operational telescopes are not considered here (0.5~\%). The selection of data
aims at efficiently accepting good weather conditions with minimal changes in
the atmosphere (mainly due to variations of the aerosol distribution). 

% Data selection cuts
The selection is based upon the following approach: Under the assumption that
the cosmic ray flux at the top of the atmosphere remains intrinsically constant
and that the bulk of triggered events are cosmic ray events, changes in the
acceptance of the detector (mainly variations of the transparency of the
atmosphere) should be easily seen in variations of the trigger rate.  For
individual data-runs (20 minutes duration) an expectation of the rate is
calculated based upon (a) the long-term average rate determined for every month
\citep{2003APh...20.267}, (b) the change of rate with altitude $20^\circ< h<60^\circ$:
$R_{cr}(h)= a_0+a_1\cdot h+a_2\cdot h^2$, $a_0=-6.363, a_1=0.556, a_2=0.004$, for $h>60^\circ$: $R_{cr}=const.=R_{cr}(60^\circ)$, and (c) a scaling according to the
number of operational telescopes ($R_5:R_4:R_3=1.16:1:0.8$).  Runs with an
average rate deviating by more than 25~\% from the expectation are rejected
(15~\% of all runs available). 
 
% Data summary
 In Table~\ref{table:data_selection}, observational data are summarized for
the individual (yearly) observational periods and for 3 intervals of zenith
angle. The total amount of data accepted sums up to 384.86~hrs of which 20~\%
were taken at large zenith angles.

%%%%%%%%%%%%%%%%%%%%%%%%%%%%%%%
\subsection{Data calibration}
%%%%%%%%%%%%%%%%%%%%%%%%%%%%%%%
\label{sect:calibration}

 The long-term calibration of data and a deep understanding of the
variations of detector performance over an extended period of time is essential
for deriving a combined energy spectrum. The calibration methods used in this
analysis are based upon \citet{2003APh...20.267} with an additional relative
inter-telescope calibration \citep{2003APh...20...1}.  The main  steps for the
calibration of image amplitude are briefly summarized below.

\paragraph{Calibration of Image Amplitude}

 In order to study the \textit{time dependence} of the telescopes' overall
efficiency $\eta$ (the ratio of the recorded amplitudes to the incident
Cherenkov light intensity) it is convenient to factorize $\eta$ into  a
constant efficiency $\eta_0$ according to the design of the telescopes at the
beginning of operation and  time dependent efficiencies $\eta_{camera}(i)$ and
$\eta_{optics}(i)$ which are unity at the beginning of operation and are
calculated for every month (period) $i$ of operation.

\begin{eqnarray}
	\eta(i) &=& \eta_0 \cdot \eta_{camera}(i) \cdot \eta_{optics}(i) 
\end{eqnarray}

 The variation of the efficiency $\eta_{optics}(i)$  is mainly caused by the
ageing of mirrors and to some smaller extent by ageing of the light funnels
installed in front of the photo-multiplier tubes (PMTs) resulting in a 
loss of reflectivity
at a steady rate of $(-3.7\pm0.3)$~\% points  per year. The variation of
$\eta_{camera}(i)$ is caused by ageing of PMTs (rate of loss $(-8.0\pm0.3)$~\%
points  per year), and occasional (typically twice per year) changes of high
voltage settings to compensate for the losses of efficiency.  The two different
efficiencies ($\eta_{camera}(i),\eta_{optics}(i)$) are disentangled by
calibrating the electronics' gain via special calibration events taken with a
laser and studying the variation of the event rate for cosmic rays.

For the calibration of the electronics, a pulsed nitrogen laser mimics
Cherenkov light flashes and allows to determine flat-fielding coefficients to
compensate for differences in the response and conversion factors for
individual PMTs.  The light is isotropized by a plastic
scintillator material in the center of the dish and illuminates the camera
directly.  The conversion factors relating the recorded digitized
pulse-height to the number of photo electrons detected in the  PMTs are derived
from the relative fluctuations in the recorded amplitudes for these 
laser-generated calibration events. 

 Finally, the efficiency of the optical system (mirrors and funnels) is
calculated indirectly: The laser calibration events are used to calculate an
average electronics gain $\eta_{camera}(i)$ for individual periods. By
comparing the average cosmic ray event rate with the expected value, an optical
efficiency $\eta_{optics}(i)$ is calculated.  
 
 This calibration method based upon the detected cosmic ray rate provides the
possibility to regularly monitor the detector efficiency  in parallel with the
normal operation of the telescopes. As a cross-check, muon runs have been
performed where muon rings are recorded by the individual telescopes with a
modified single-telescope trigger. The muon rings provide an independent and
well-understood approach to determine the efficiency of the telescope
\citep{APh1994...2..1}.  The absolute value of the efficiency determined from
the muon events is $\approx 11$~\% which is very close to the design
specification of $12$~\% used for the simulations.  The muon events were taken
at regular intervals and verify the decay of the reflectivity of the mirrors
and the ageing of the PMTs derived from the cosmic ray rate.  Further details
on the procedure of calibration of this data-set are given in
\citet{2003APh...20.267}.

%Since the read-out system is based upon 120 MHz FADC digitization electronics,
%the arrival time of Cherenkov pulses is recorded in the individual pixels at a
%resolution better than 1~nsec and is used to reconstruct the time profile of
%the shower images \citep{1999APh....11..363}.  Again, the laser-generated
%calibration events (which are assumed to be isochronous and homogenous) are used
%to measure the difference in signal travel times in the individual channels.
%However, the arrival time information does not substantially improve the
%separation of gamma and hadron induced events with respect to the separation
%obtained by imaging methods and is not used for this analysis.

\paragraph{Inter-telescope calibration}

 The multi-telescope view of individual events opens the possibility for
inter-telescope calibration. The method applied here follows
\citet{2003APh...20...1} where for pairs of telescopes with a similar distance
to the shower core, relative changes in the response are calculated. For the
procedure, the data-set was separated in 5 observational seasons coinciding
with the yearly visibility period of the Crab nebula. Events from an extended
region in the camera (within 0.5$^\circ$ radial distance to the center of the
camera) passing the image shape selection criteria for $\gamma$ events have
been selected.  In principle, the calibration can be performed using
exclusively $\gamma$-rays.  The caveat of selecting only $gamma$-events is the
comparably small number of events which limits the statistical accuracy.  By
choosing background events from a large solid angle, the event statistics is
sufficient to calculate the relative calibration factors within 1-2\,\%
relative accuracy.  Various systematic checks on simulated data and different
$\gamma$-ray sources have been performed to verify the systematic accuracy of
the procedure to be within $2~\%$.

%%%%%%%%%%%%%%%%%%%%%%%%%%%%%%%%%%%%%%%%%%%%%%%%%%%
\subsection{Data reduction and event reconstruction}
%%%%%%%%%%%%%%%%%%%%%%%%%%%%%%%%%%%%%%%%%%%%%%%%%%%
\label{section:data_reduction}
\paragraph{Image cleaning}

 The data are screened for defective channels (on average 1~\% of the pixels are
defective) which are subsequently excluded from the image analysis. Pixels
exposed to bright star light are excluded from the trigger during data-taking
and are subsequently not used for the image analysis. After pedestal
subtraction and calibration of the amplitudes (see above), a static two stage
tail-cut is applied, removing in the first step all pixels with amplitudes of
less than 3~p.e.  and in a second step removing all pixels below 6~p.e. unless
a neighboring pixel has registered an amplitude exceeding 6~p.e.

The tail-cut value is sufficiently high to ensure that the image is not
contaminated by night sky background induced noise. The typical RMS noise level
varies between $\approx 0.8$~p.e. and $\approx 1.5$~p.e.  Cleaned images
exceeding a size of 40~p.e., and with a distance of the center of gravity of
less than 1.7$^\circ$ from the camera center are combined in the stereoscopic analysis  which is used
to reconstruct the shower parameters. 

\paragraph{Image and Event selection}
 Selection of images and events are optimized separately for the three
different studies carried out (spectrum, morphology, and pulsation).  The
selection applied for the energy spectrum will be explained below. A summary of
the different cuts applied for the different analyses is given in
Table~\ref{table:selection}.  

The cuts imposed on the reconstructed events to
be used to determine the energy spectrum are chosen to minimize the systematic
and statistical uncertainties while retaining enough sensitivity to reconstruct
the energy spectrum at low energies (sub threshold) and at high energies. In
order to keep systematic uncertainties small, a loose angular cut is chosen
(i.e. substantially larger than the angular resolution) on the half-angle of
the event direction with respect to the source direction: Events with a
half-angle $\vartheta<0.225^\circ$ are retained.  

The separation of $\gamma$ and hadron induced events is made by applying a cut
on a parameter related to the width of the image: mean-scaled width ($\tilde
w$). This parameter is calculated for each event by scaling the width of
individual images to the expectation value for a given impact distance of the
telescope to the shower core,  the size of the image, and the zenith angle of
the event \citep{1997APh...8...1}. The individually scaled width values are
averaged over all available images. For $\gamma$-ray like events this value
follows a Gaussian distribution centered on an average of $1$ with a width of
$0.1$. For this analysis, we choose a cut on $\tilde w<1.1$ with an efficiency
of 85~\% for $\gamma$-ray events. The efficiency for $\gamma$-events changes
very little with energy and for different zenith angles. 

 Furthermore, the core position is required to be within 200~m distance of the
central telescope. This cut is relaxed for events with a zenith angle of more
than 45$^\circ$ to 400~m to benefit from the increased collection area for
larger zenith angle events \citep{1987JPhysG...13...533,2002JPhysG...28..2755}.
Finally, for the stereoscopic reconstruction, a minimum of three useful images
and a minimum angle of $20^\circ$ ($10^\circ$ for large zenith angles) subtended
by the major axes of at least one of the pairs of images is required.

\paragraph{Event reconstruction}

The geometric reconstruction of multiple images of the same shower allows to
unambiguously reconstruct \textit{event-by-event} the direction of the shower
axis with a resolution of the angular distance of $0.06^\circ=3.6$\arcmin\
for
events with three or more images.  The impact point of the shower is
reconstructed  in a similar way and for a point source with known coordinates,
the resolution for the shower impact point is better than 5~m
\citep{1999APh....12..135H,2000A&A...361.1073A}.

 The stereoscopic observation method allows to reconstruct  the
position of the maximum of the longitudinal shower development 
\citep{2000ApJ...543L..39A} which is used to
estimate the energy of the primary particle \citep{2000APh....12..207H}.

Whereas for single-telescope observations the  reconstruction of the primary
energy is influenced by the strong fluctuation of the longitudinal shower
development and the subsequent variation of the measured light density in the
observer's plane, the energy resolution for stereoscopic observations is
improved by correcting for the fluctuation of the shower maximum. The relative
energy resolution $\sigma_E/E$ typically reaches values below 10~\% for
energies above the threshold and moderately increases to $15$~\% at threshold
energies. Note that the bias of systematically over-predicting the energy at
threshold is reduced considerably (conventional energy reconstruction produces
a $\approx 40~\%$ bias whereas the improved method reduces this bias to $\approx 5~\%$).

%%%%%%%%%%%%%%%%%%%%%%%%%%%%%%%%%%%%%%%%%%%%%%%%%%%
\subsection{Energy spectrum}
%%%%%%%%%%%%%%%%%%%%%%%%%%%%%%%%%%%%%%%%%%%%%%%%%%%
\label{section:energy_spectrum}

   The method of reconstructing the energy spectrum follows the approach
described in \citet{1998APh...9..1,1999A&A...349...11A}. The method is
considerably extended to deal with the temporal variations of the detector over
the course of its lifetime \citep{2002A&A...393...89A} and to improve
on the extraction of weak signals \citep{2003A&A...403L..523}.  The collection areas
$A_{eff}(\mathrm{E}_\mathrm{rec},\theta,i,n_{tel})$ are derived from simulated
air showers \citep{2000NIMA...450..419} subject to a detailed detector
simulation \citep{Hemberger}. The same analysis chain as for the data  is applied to simulated events.  

The collection areas for 5 discrete zenith angles (0$^\circ$, 20$^\circ$,
30$^\circ$, 45$^\circ$, 60$^\circ$) are calculated as a function of
reconstructed energy between 0.1 and 100~TeV.  Moreover, the detector
simulation is repeated for the 68 individual observation periods $i$ (see Section \ref{sect:calibration}) using the
temporal changes in efficiency as described above to model the variation of the
response of the detector. For each period and zenith angle, three individual
simulations are performed with 3, 4, and 5 operational telescopes.  In total,
1020 different collection areas are generated (5 zenith angles, 3 telescope
set-ups, 68 periods) and stored in a database. 

 For each individual event that is reconstructed and passes the event-selection
criteria, a collection area is calculated. The procedure to interpolate the
collection area for arbitrary zenith angles is described in detail in
\citet{1999A&A...349...11A}. The events are counted in discrete energy bins and
for each bin $j$ covering the interval $E,E+\Delta E$ the differential flux is
calculated: 

\begin{eqnarray}
\frac{d\Phi_j}{dE} &=& (T\,\Delta E)^{-1}\cdot
\left(\sum \left(A_{eff}^{ON}\right)^{-1}-
                   \alpha \sum \left(A_{eff}^{OFF}\right)^{-1}\right)
\end{eqnarray}

 The factor $\alpha$ is given by the ratio of solid angles of the ON and OFF
observation regions respectively: 
$\alpha=\Omega_{OFF}/\Omega_{ON}$.  For the observations considered here,
five separate OFF regions with equal distance to the camera center are
chosen ($\alpha=0.2$). 

For simplicity, the collection area above 30~TeV is assumed to be constant and
equal to the geometrical size ($\pi\cdot200^2$~m$^2$ for zenith angles smaller
than 45$^\circ$, $\pi\cdot400^2$~m$^2$ for zenith angles beyond 45$^\circ$). 
{This has been checked with simulated air showers as shown in \citet{1999A&A...349...11A}.}

 The effective observation time  $T$ is calculated from the sum of time-differences of events.  A
correction for the dead-time (2~\%) is applied and occasional gaps in
data-taking (less than 0.5~\%) due to network problems are excluded from the
effective observation time. 
 
\begin{figure} 
\plotone{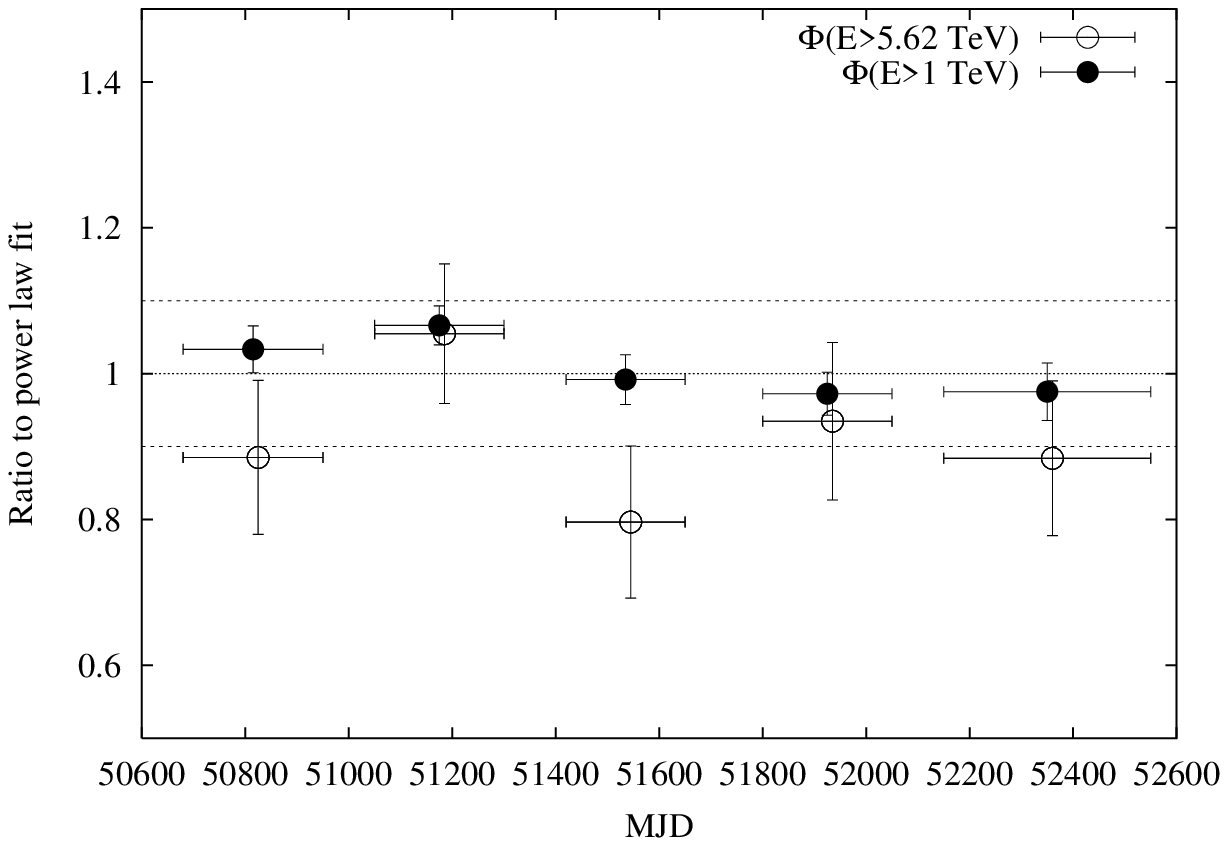}
\caption{\label{fig:int} As a check for unnoticed temporal changes in the
instrumental response, the integrated fluxes for the five observing years are
compared with each other. The respective values are normalized to the integral
flux derived from the power-law fit to the entire data-set. As is clearly seen
from the relative deviations of $\Phi(E>1~\mathrm{TeV})$ systematic effects are
smaller than 10~\% indicated by the dashed lines which translates into a
relative uncertainty  of the absolute energy calibration of better than 6~\%
(assuming a power law type spectrum with an index of -2.62). The integrated
fluxes above 5.62~TeV show larger fluctuations which are still consistent with
a constant flux.} \end{figure}

 The resulting integrated fluxes calculated individually for the 5 different
observing seasons show consistent values with possible systematic differences
smaller than 10~\% (see Fig.~\ref{fig:int}). 
The integrated
fluxes above 1 and 5.6~TeV shown in Fig.~\ref{fig:int} are normalized to the
respective integral flux calculated from the power law fit to the entire
spectrum. 
Taking the 
maximum deviation as an estimate of systematic
uncertainty this translates into a possible variation of the absolute energy
calibration of the system from year to year of less than 6~\%. 
 As is already evident from Figure~\ref{fig:int} the integrated flux above
5.6~TeV is on average slightly lower than the power law fit expectation.  This
is an indication for a steepening of the energy spectrum towards higher
energies which will be further discussed in Sect.~\ref{section:model}.  At this
point, the variations of the integral fluxes are consistent with systematic
effects and no temporal variation of the source can be claimed. 

For the given constant energy resolution over a broad range of energies
systematic effects of bin to bin migration in the presence of a steeply falling
spectrum are not important.  
With the use of
an energy reconstruction method which takes the height of the shower maximum
into account (see Section~\ref{section:data_reduction}), 
systematic bias effects near threshold are greatly reduced. In
any case, the method applied automatically takes the effect of a small bias in
the energy reconstruction into account.

The bin width in energy is chosen to reduce the interdependence of the bins ($\Delta E=
3~\sigma_E$, $\sigma_E$ is the typical energy resolution $\sigma_E/E\approx 10$~\%) and therefore simple $\chi^2$ minimization methods can be applied
to fit arbitrary functions to the data.  The errors on the parameters of the
fit are calculated with the MINUIT package routines \citep{Minuit} and have
been tested with simple Monte Carlo type simulations of random
experiments.  Bins with a signal
exceeding 2 standard deviations according to Eqn.~17 in
\citet{1983ApJ...272..317L} have been included in the calculation of
differential fluxes. Above 10~TeV, the bin width has been increased to
compensate for the fast decrease of event statistics. 

\subsection{Position and size of emission region}

\label{section:morphology}
% The arrival directions for the events are calculated employing 
%an advanced algorithm for stereoscopic reconstruction 
%(Algorithm \#2 in \citet{1999APh....12..135H}).  
%Besides improving the accuracy for the
%angular reconstruction, an uncertainty on the reconstructed direction
%is estimated $\sigma_{xy}$ which can be used to 
%select a sample of well-reconstructed events. 

% Two different studies are carried out to constrain the emission region: Firstly, the position of the emission region is
%constrained for different bins of energy. Secondly, the size of the emission region is
%constrained.  

% However, the selection of events is slightly different for the two cases. The cut on 
%the estimated error $\sigma_{xy}$ of the arrival direction is chosen to be very tight ($\sigma_{xy}<0.05^\circ$)
%to improve the sensitivity for the size of the emission region. For locating the emission region, a
%slightly looser cut on $\sigma_{xy}<0.07^\circ$ has been chosen. 

In order to determine the position and size of the emission region,  
a 2-dimensional Gaussian function (Eqn. \ref{eqn:2d}) is used to fit a
histogram containing the arrival directions in discrete bins of $0.05^\circ$
side length in a projection of the sky. The bins are uncorrelated and chosen 
as a compromise between the systematic effect of discrete bins and sufficient
statistics in the bins.

\begin{eqnarray}
\label{eqn:2d}
 f(x,y)&=&f_0\cdot \exp\left[-\frac{(x-\langle x\rangle)^2}{2\,\sigma_x^2}
-\frac{(y-\langle y\rangle)^2}{2\,\sigma_y^2}\right]
+f_{ped}
\end{eqnarray}

 The coordinates ($x,y$) chosen are sky coordinates in declination ($\delta$)
and right ascension ($r.a.$). The declination is measured in degrees whereas
right ascension is in units of hours.
The constant pedestal $f_{ped}$ is calculated 
from averaging the counts/bin in an annulus around the source region with 
inner radius $0.2^\circ$ and outer radius $0.35^\circ$.  

 The width of the distribution $\sigma_0=\sqrt{\sigma_x^2+\sigma_y^2}$ is
assumed to be symmetric ($\sigma_x=\sigma_y$) and a convolution of the
instrumental point spread function ($\sigma_{psf}$) and a source size
($\sigma_{src}$): $\sigma_0=\sqrt{\sigma_{psf}^2+\sigma_{src}^2}$. This is a
simplification assuming that the point spread function and the source size are
both following a Gaussian distribution \citep{2000A&A...361.1073A}. 

 The analyses proceeds in two steps: In the first fit, a loose cut on the 
event selection is applied rejecting events with an estimated
angular resolution $\sigma_{xy}>0.10^\circ$ and the source extension is set to
zero ($\sigma_0=\sigma_{psf}$). In the second fit, the event selection is rejecting
events with an estimated angular resolution $\sigma_{xy}>0.06^\circ$ and the source
position is kept at the values found in the first fit whereas $\sigma_{src}$ is left
as a free parameter. 

\begin{figure}
\plotone{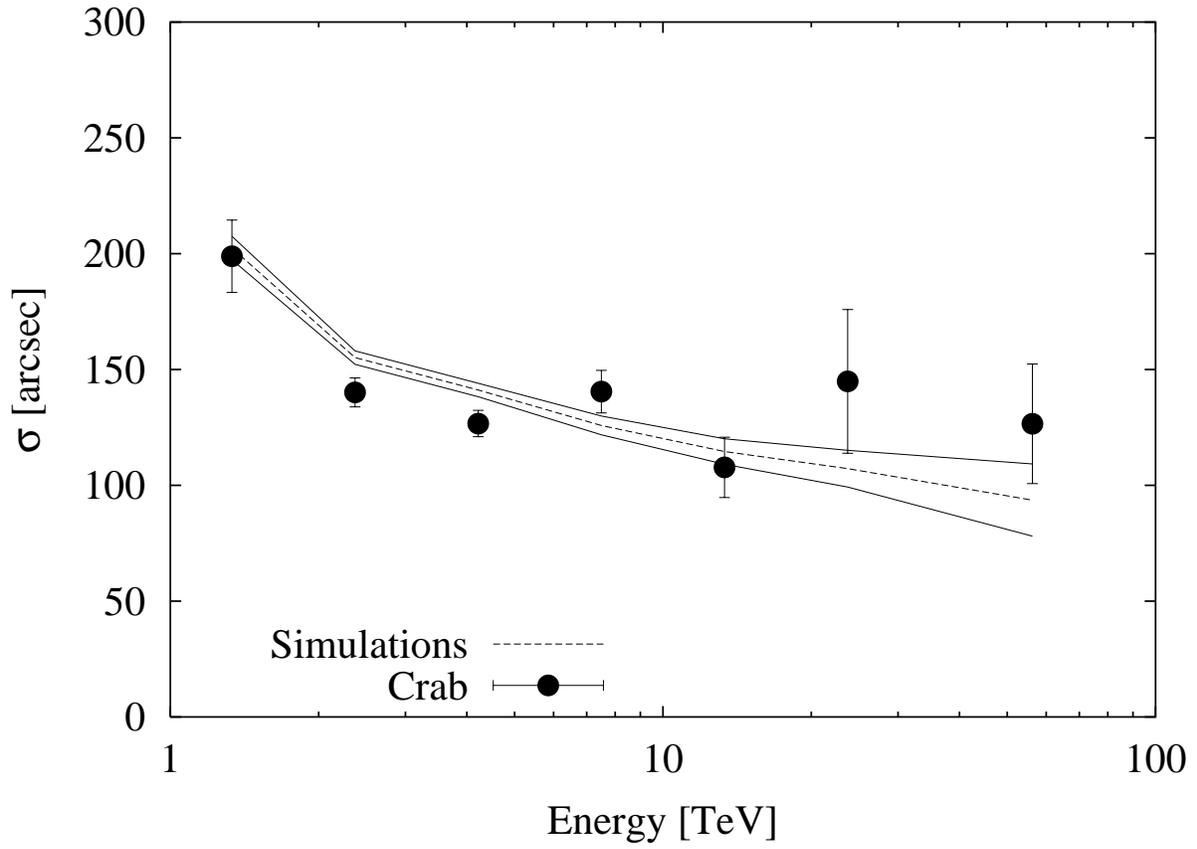}
\caption{\label{fig:ang} As a function of reconstructed energy, the size of the point spread function 
observed from the Crab nebula does not differ from the predicted values from simulation (dashed and solid
lines indicate average and 1$~\sigma$ uncertainty.) }
\end{figure}

The instrumental point spread  function $\sigma_{psf}$ is characterized using the predicted
value of Monte Carlo simulations  that have been checked against the performance
for extra-galactic objects like Mkn~421, Mkn~501, and 1ES1959+650 \citep{2003A&A...406L....9}. After
applying a cut 
on the estimated error $\sigma_{xy}<0.10^\circ$ selecting 75~\%
of the gamma-ray events, the point-spread function is well-described by a Gaussian function.
The point spread function weakly depends on the energy and the
zenith angle after applying this selection cut (see also Fig.~\ref{fig:ang}).

 Applying a dedicated analysis for high energy events (e.g. raising the tail
cut values) improves the angular resolution at the high energy end, but
requires reprocessing of all raw events. Given that systematic uncertainties of
the pointing of the instrument (25\arcsec) and the low photon numbers at high
energies strongly diminish the sensitivity  for the source location and
extension, a
substantial improvement is not expected. 
%With the source position fixed, the only
%free parameters left in
%Eqn.~\ref{eqn:2d}  are the normalization and $\sigma_0$.  

The size of the excess region is determined by fitting a Gaussian function (see
Eqn.~\ref{eqn:2d}) with fixed position $\langle x\rangle, \langle y\rangle$ 
to a histogram with the reconstructed arrival directions.
As a compromise between diminishing statistics and small intrinsic point spread
function, a cut on $\sigma_{xy}<0.06^\circ$ has been chosen which accepts 25~\%
of the gamma-ray events and 3\% of the background events.  In
Fig.~\ref{fig:ang}, the resulting size of the point spread function is
indicated together with the prediction from simulation. In order to predict the
size of the point spread function, the simulated air showers have been weighted
according to the distribution of zenith angles and energies found in data. 
The procedure has been checked
against the point spread function of Mkn~421 and Mkn~501 up to energies of
10~TeV.  Based upon the predictions and measurements with their respective
statistical uncertainties,  upper limits on the source size are calculated
with a one sided confidence level of 99.865~\% ($3~\sigma$). 
A systematic uncertainty of
25\arcsec\ has been added to the upper limits motivated by the absolute pointing
accuracy. This can be considered a conservative estimate.

\subsection{Phase resolved analysis}
\label{section:phase}

 For events recorded after spring 1997, a global positioning system (GPS) time
 stamp is available for each individual event.  With the high accuracy and
 stable timing of individual events, a phase resolved study of the arrival
 times coherently over extended observation time is feasible.

In order to search for a pulsed signal from a pulsar which is not a part of a
binary system, it is sufficient to use the arrival time at the barycentre of
the solar system and to calculate a relative phase with respect to the arrival
time of the main pulse as seen in radio observations. 

The conversion of the GPS time stamp to the solar barycentre 
is done using the direction of the pulsar and  the
 JPL DE200 solar system ephemeris \citep{1982A&A...114..297S}
which have a predicted accuracy of 200~m confirmed 
by the updated DE405 ephemeris \citep{StandishMemo,2001CeMDA..80..249P}.

The algorithm applied includes a higher order correction for a relativistic
effect that occurs whenever photons pass close to the sun and are slightly
deflected in the gravitational field \citep{Shapiro}. The resulting delay
however is for the night-time observations of Cherenkov telescopes in any case
negligibly small. 

The Crab pulsar ephemerides are taken from the public Jodrell Bank database
\citep{Jodrell} and are used to calculate the relative phase of each event. A
linear order Taylor expansion is used to calculate the arrival time of pulses
for an arbitrary time between two radio measurements. For this purpose the
derivative $\dot P$ of the period $P$ is used  in the expansion of the
ephemerides. In the absence of glitches, this method gives accurate arrival
times with an accuracy of 250~$\mu$s. Whenever a glitch occurs with an abrupt
change of $P$, a new interpolation period is started taking the change in $P$
and $\dot P$ into account.

 The timing analysis has been checked with optical observations of the Crab
 pulsar using the prototype stand-alone Cherenkov telescope
 \citep{ICRC...OG2.2...2449}.  Additionally, data taken with the H.E.S.S.
 instrument on the optical Crab pulses \citep{ICRC...OG2.5...2897}  have been
 used to verify the procedure of the solar barycentric calculation. 

 Subsequent phase folding of the HEGRA (ON-source and OFF-source) data results
 in a pulse profile.  The pulse profile is split into bins according to the
 pulse shape seen by the EGRET instrument \citep{1998ApJ...494..734F} using
 bins for the main pulse (P1) with leading and  trailing wing of the main pulse
 (LW1, TW1), bridge (B), secondary pulse (P2) with  trailing and leading wing
 (LW2, TW2), and off-pulse. 

 A Pearson-$\chi^2$-test on a uniform distribution
has some arbitrariness concerning the way that the phase distribution is 
discretized in bins.
In addition to this bias, the sensitivity of the $\chi^2$-test is inferior to
tests invoking the
relative phase information of successive events like the Rayleigh-
\citep{Mardia}, $Z_2^m$-\citep{1983A&A...128...245}, and $H_2$-test
\citep{1989A&A...221..180}. Here, a $Z_2^2$-test has been applied 
which results in an
improved sensitivity with respect to the Rayleigh- or $\chi^2$-test statistics
for a sharp bimodal pulse form similar to the one seen by EGRET for the Crab
pulsar.%  For the sake of completeness, a Rayleigh test is also performed. 

The tests for periodicity are performed on seven
energy intervals covering the energy from 0.32 up to 100~TeV.  
This procedure is motivated by the predictions of sharp
features in the spectrum of the pulsed emission \citep{2001ApJ...558..216H}.

\section{Results and interpretation}

\subsection{Energy spectrum}
\label{subsec:Energy_spectrum}
\begin{figure} 
\plotone{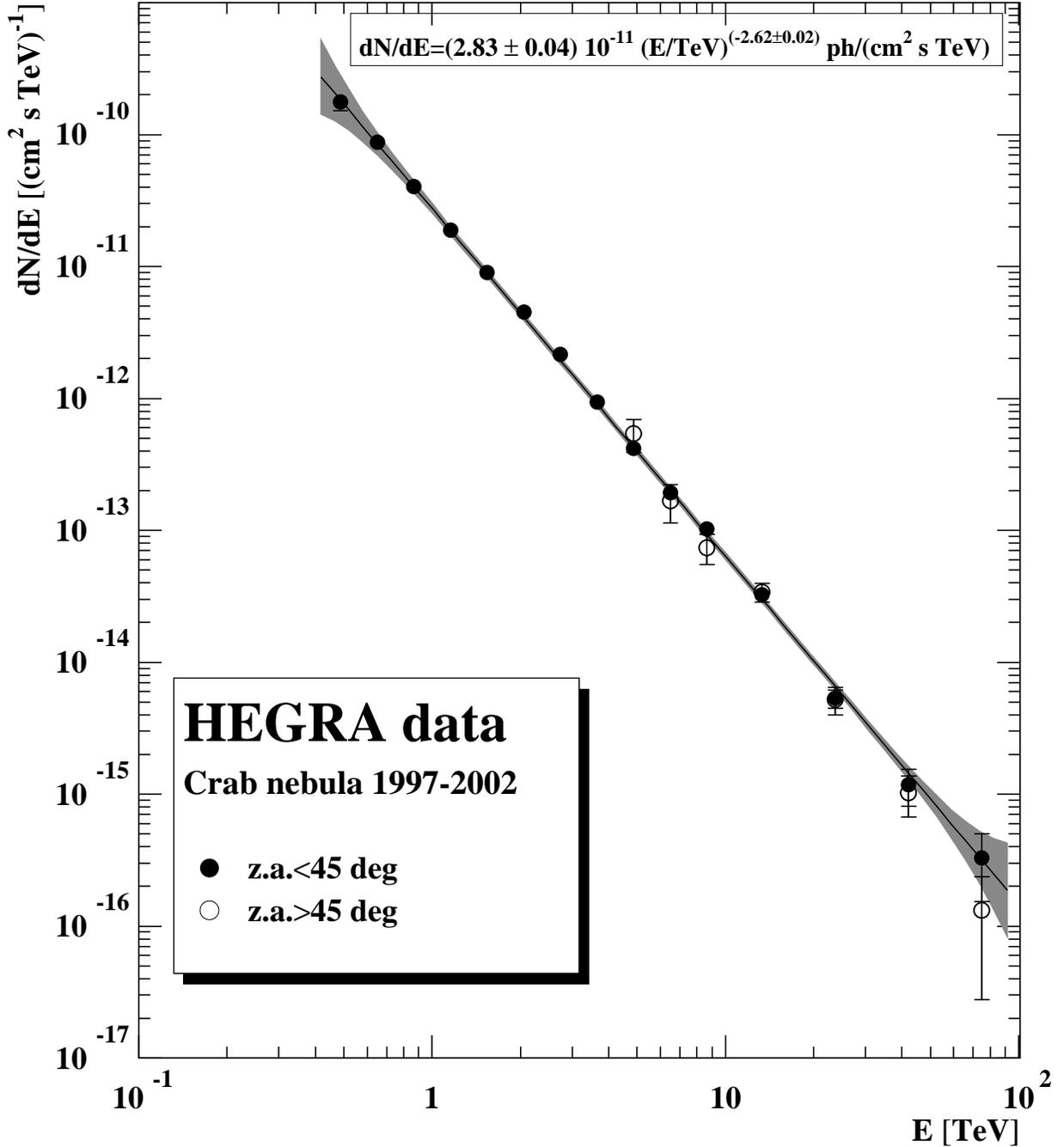}
\caption{\label{fig:spec1} As a result of the spectral analysis in two
different zenith angle intervals
(6.2$^\circ$-45$^\circ$,45$^\circ$-65$^\circ$), the differential energy
spectrum covers the range from 0.5~TeV up to 100~TeV. The significance of the
data  point centered on 86~TeV is 2.7~$\sigma$ after combining the events from
all zenith angles. The grey shaded region indicates the range of systematic
errors which are most prominent in the threshold region where the statistical
errors are smaller than the systematic uncertainties.  } \end{figure}

 The reconstructed differential energy spectra for two different ranges in zenith
 angle are shown in Fig.~\ref{fig:spec1}.  Systematic uncertainties are
 estimated based upon possible variations in the threshold region unaccounted
 for in the simulations, uncertainties in the nonlinear response of the PMTs
 and read-out chain at high signal amplitudes. The range of systematic
 uncertainties is indicated by the grey shaded region in Fig.~\ref{fig:spec1}. 
 In addition to the energy dependent uncertainties shown by the 
 grey shaded region, the global energy scale is 
 uncertain within 15~\%.

 The dominating uncertainties are
 visible in the threshold region above 500~GeV. The estimate of the systematic
 uncertainties is shown to be quite conservative judging from the smooth connection
 of the flux measured at small zenith angles with the flux measured at larger
 zenith angles at $\approx 6$~TeV. The flux measured at the threshold for the
 data set with zenith angles larger than $45^\circ$ is in very good agreement
 with the measurement from smaller zenith angles. 

Additionally, the good agreement of the two different energy spectra at very high energies shows that the
systematic effect of the nonlinear response at high signal amplitudes is
probably smaller than estimated. The nonlinear response affects mainly the
small zenith angle observations where for a given energy of the shower the
average image amplitude is much higher (by a factor of $4\ldots6$) than for
larger zenith angles. The effect of the saturation/nonlinearity is expected to
be negligible at larger zenith angles. The good agreement shows that the
correction applied to the high signal amplitudes is accurate.

 For the purpose of combining data taken at different zenith angles, the same
 approach as described in \citet{2000ApJ...539..317A} is followed. The combined
 energy spectrum (Table~\ref{table:spec}) is well approximated by a pure power law of the form $d\Phi/dE=\Phi_0
 \cdot (E/\mathrm{TeV})^{\Gamma}$ with $\Phi_0=(2.83\pm0.04_\mathrm{stat}\pm0.6_\mathrm{sys})$~\dflux and
 $\Gamma=-2.62\pm0.02_\mathrm{stat}\pm0.05_\mathrm{sys}$. The $\chi^2_{red}(d.o.f.)=1.3(13)$ indicates that 
 deviations from a power law do not appear to be very significant. The systematic
 errors quoted on the parameters are the result of varying the data points within 
 the systematic errors (the grey shaded band in Fig.~\ref{fig:spec1}) and 
 for the normalization $\Phi_0$ the uncertainty of the energy scale is included.

 Even though a power law fit to the data is a good approximation without
 statistically significant deviations,  
 a systematic deviation from a power law  in the form of a steepening 
can be seen upon closer inspection (see also Fig.~\ref{fig:diff}) 
and will be discussed in Section~\ref{section:Comparison}.

\subsection{Source position and morphology}
\label{section:morphology_results}
In a previous HEGRA paper \citep{2000A&A...361.1073A}, the emission size region of the Crab nebula for energies
up to 5~TeV has been constrained with a smaller data-set (about one third of
the data used here) to be less than 1.7~arc~min .  

 Here, results from a similar analysis technique are presented. The most
noticeable difference with respect to the previously published results is that
the upper limits are calculated within 7 independent energy bins covering the
energy range from 1 to 80 TeV. In principle, an ionic component in the wind
could be a rather narrow feature in the energy spectrum and therefore might
have gone unnoticed in an analysis using the integral distribution of all
gamma-ray events.

Taking all data, the center of the emission region is determined to be 
r.a.=$5^h34^m(31.1\pm0.2_\mathrm{stat}\pm1.8_\mathrm{sys})^s$, 
$\delta=22^\circ0'(52\pm3_\mathrm{stat}\pm25_\mathrm{sys})''$ (J2000) which is
shifted by $(12.4\pm3.1_\mathrm{stat}\pm25_\mathrm{sys})$\arcsec\ angular distance to the west of the
nominal position of the pulsar (J2000.0)
r.a.=$5^h34^m31.97^s$, $\delta=22^\circ0'52.1''$ \citep{1999A&AS..136..571H}. This shift
is consistent with the expectation based upon the centroid position of the
X-ray emitting nebula. The centroid position of the X-ray emitting nebula
 derived from public Chandra 
data of the Crab  excluding the pulsar emission
is shifted by 9\arcsec\ to the west. In declination the
position of the TeV emission region is consistent  with the position of the pulsar:
 ($1.9\pm3.0_\mathrm{stat}\pm25_\mathrm{sys}$)\arcsec\ angular
distance. The 
centroid position of the X-ray emitting nebula as measured with the
Chandra X-ray telescope is shifted to the north by 13\arcsec. 

\begin{figure}
\plotone{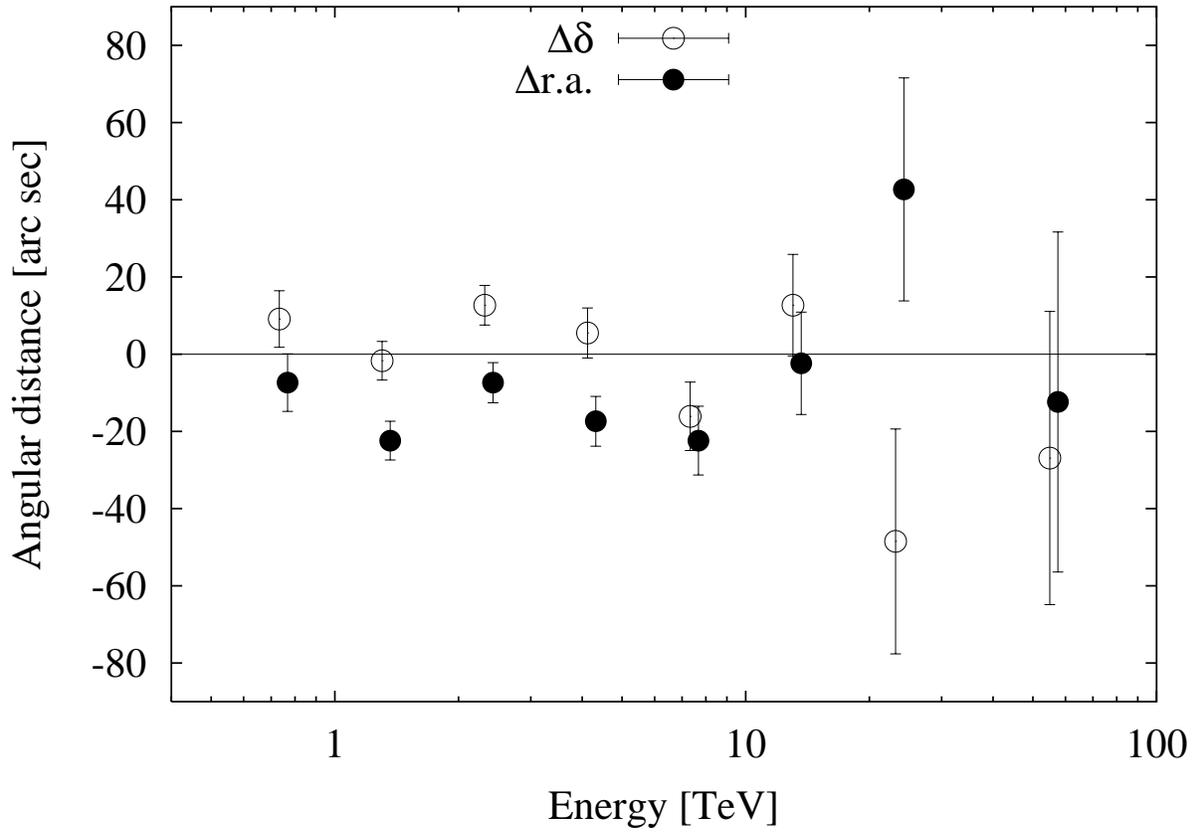}
\caption{\label{fig:dev} As a function of reconstructed energy, the position of the emission region 
does not show significant variation. The open symbols indicate the deviation of the reconstructed 
declination to the declination of the pulsar ($\Delta \delta$), filled symbols indicate the deviation of the
right ascension ($\Delta$ r.a.).}
\end{figure}

Checks on the data split into the
yearly observation campaigns confirm the offset to be present throughout the
observation time. The statistical uncertainty  on the position is smaller than
the observed shift in right ascension. However, given the systematic uncertainty of 
25\arcsec\ derived from the pointing calibration, it is not possible to identify
the position of the emission region with the pulsar or with structures in the
nebula. The shift remains constant for different energy bins as shown in
Fig.~\ref{fig:dev}.

\begin{figure}
\plotone{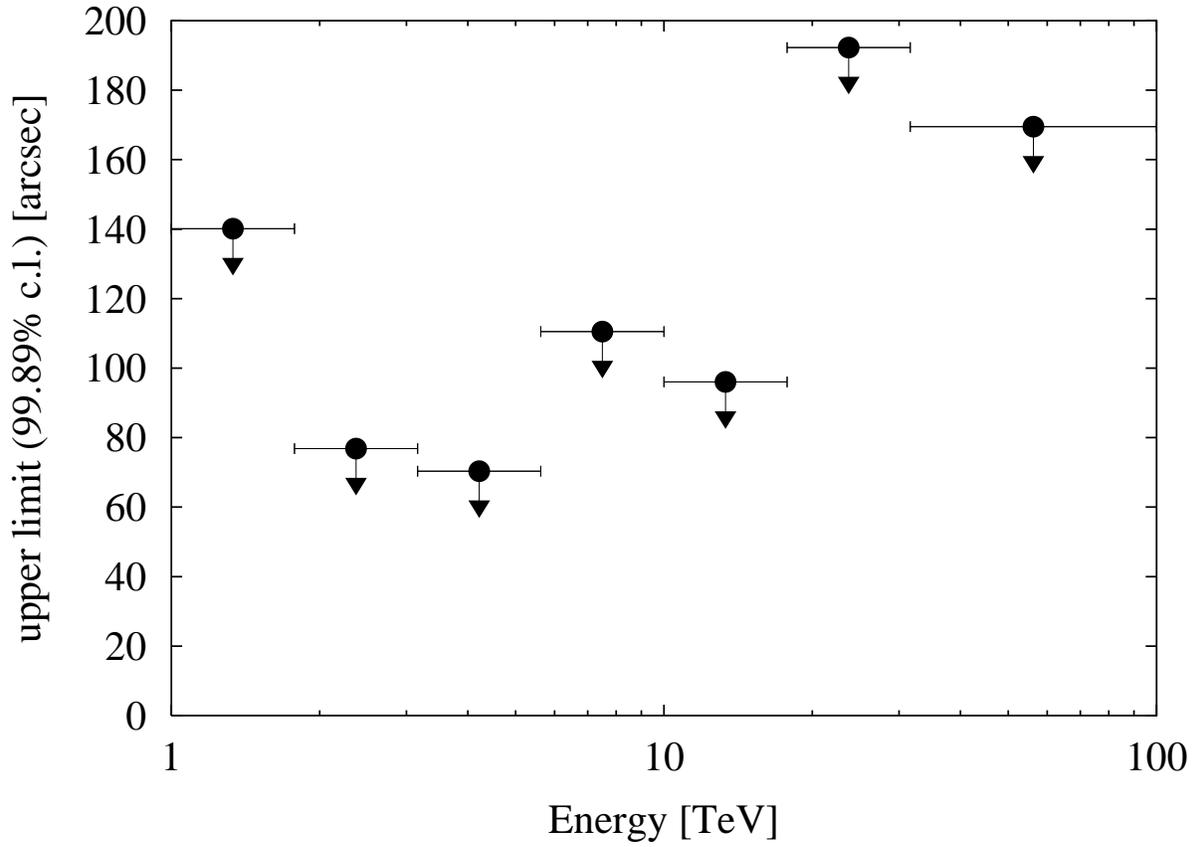}
\caption{\label{fig:ul} Based upon the predicted point-spread function from simulation 3$\sigma$ upper limits
for a Gaussian type source size have been calculated. These are independent (differential) upper limits improving
sensitivity for a particular narrow (in energy) feature expected from an ionic component in the wind.}
\end{figure}

The limits on the source extension  $\sigma_{src}$ (with a confidence level of 99.865~\%)
are shown in Fig.~\ref{fig:ul}. Clearly, a source size
exceeding  2\arcmin\ can be excluded at energies below 10~TeV. At higher
energies, the limit given here 
constrains the size to be less than 3\arcmin.  The expected
source size in the leptonic model would be close to 20\arcsec\ whereas for an
ionic component the size of the emission region would 
exceed 3\arcmin. Given the upper limits
here, a narrow (in energy) emission component as predicted by
\citet{2003A&A...402..827A,2003A&A...405..689B} which could lead to an increase
of the source size for energies where this emission component dominates, is not
found. This is consistent with the upper limit derived on the fraction
of the spin-down luminosity present in kinetic energy of ions in the wind (see 
Section~\ref{section:model}).

\begin{figure}
\plotone{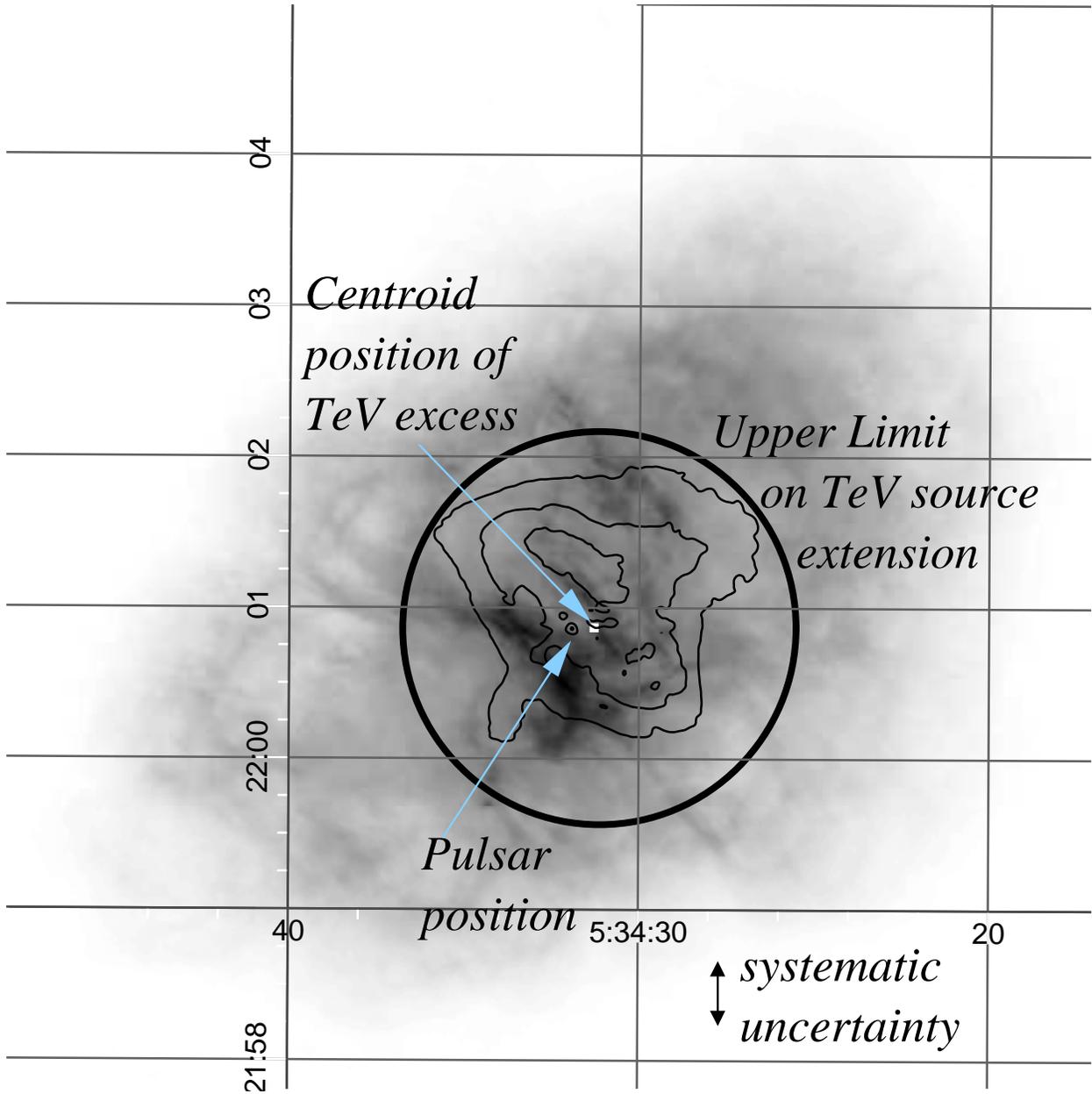}
\caption{Shown in greyscale is the radio map of the Crab nebula combined with
X-ray contours. The solid line circle indicates the upper limit on the
\label{plot:crab_map}
TeV source size and as a white square the position of the TeV centroid 
(the sidelength indicate the 3\arcsec\ statistical error).}
\end{figure}

  In Fig.~\ref{plot:crab_map} the upper limit on the source extension between
3 and 5.6~TeV is compared with the radio (greyscale) and X-ray map (contours)
taken from the Chandra Supernova Remnant Catalog\footnote{See http://snrcat.cfa.harvard.edu}. The solid circle indicates the upper limit on the TeV source
size.  The small white square depicts the position of the 
TeV emission centroid where the sidelength of the square is the statistical 
uncertainty on the position.

\subsection{Phase resolved analysis}
\label{section:phase_result}

\begin{figure}
\plotone{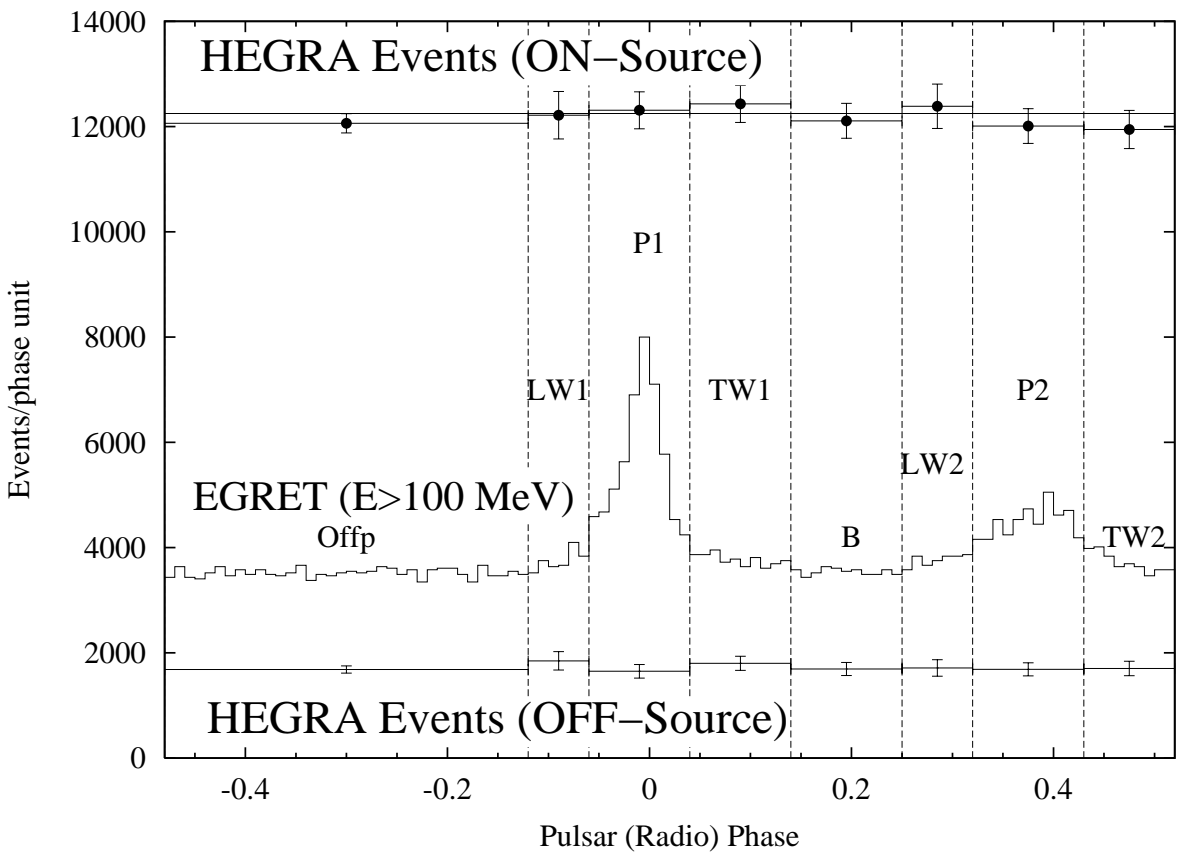}
\caption{\label{fig:crab_pulse} The phasogram of the Crab emission at two different energies: EGRET measurements
show above 100~MeV a bimodal distribution. The different characteristic pulse intervals are marked (see text for
further details). For the entire HEGRA data-set  the ON-source and OFF-source distribution in the same phase bins
is shown without evidence for a signal.
}
\end{figure}

 The phasogram obtained by folding the events registered from the direction
of the Crab pulsar and a background region are shown in Fig.~\ref{fig:crab_pulse}
together with the phasogram obtained with EGRET at lower energies \citep{1998ApJ...494..734F}. 
 Additionally, the search for pulsed emission is carried out on short data-sets
 covering a month each. This is a test to check whether unnoticed long term
 instability effects of the GPS timing information might have smeared a signal
 after combining data taken over long periods of time. The distribution of
 probabilities derived from the various tests (Rayleigh and $Z_2^2$) is
 consistent with being uniform for ON and OFF source data. No episodic excess
 is observed.

 In the absence of a pulsed signal, upper limits are calculated.  The strong
 persistent signal of the Crab nebula is causing a substantial background to
 a possible pulsed signal and needs to be considered for the calculation of upper
 limits.

The method described in \citet{1999A&A...346...913} is used to constrain a
possible pulsed fraction of the unpulsed flux. Here, we use the main pulse (P1)
region between the phase $-0.06$ and $0.04$  to derive upper limits at
the confidence level of 99.865~\,\% ($3~\sigma$).

\begin{figure}
\plotone{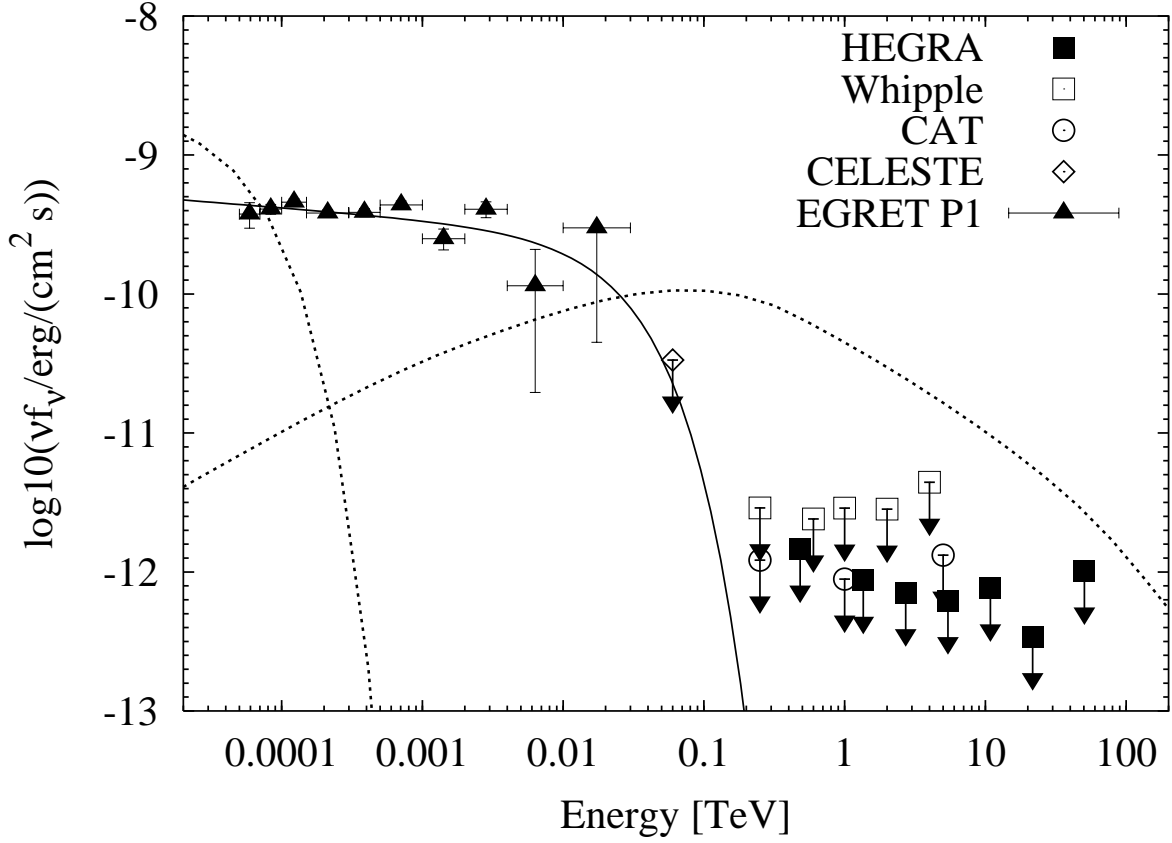}
\caption{\label{fig:crab_pulse_spec} 
Result of a search for pulsed gamma-ray emission with HEGRA (upper limits indicated by filled squares). The integral upper limits
quoted for the CAT, CELESTE, and Whipple group have been converted by assuming a differential spectral index of -2.4 at energies 
below 1~TeV and -2,6 above 1~TeV.  For comparison, the energy spectrum for the main pulse as detected by the EGRET
spark chamer is shown by filled triangles.
 The existence of a cut-off is expected in most theoretical models
which propose the origin of the Gamma-rays to be close to the polar cap or the magnetosphere (outer gap). The 
upper limits shown are consistent with the picture of a cut-off at a few ten GeV. 
The solid line is 
fit of a power-law function with photon index $\Gamma=-2.08$ and an exponential cut-off at 25~GeV to the 
EGRET data and the CELESTE upper-limit as suggested in \citet{Durand}. The dotted line are the synchrotron and inverse Compton emission of the
nebula as given by the model prediction explained in the text.}
\end{figure}

 The upper limits for 7 bins of energy are summarized in
Table~\ref{table:pulsed_limits} and shown in Fig.~\ref{fig:crab_pulse_spec}
together with results from EGRET \citep{1998ApJ...494..734F} and ground-based gamma-ray detectors: CAT \citep{musquere}, Whipple
\citep{2000ApJ...531..942L}, and CELESTE \citep{2002ApJ...566..343D}.  The HEGRA
upper limits are marginally 
dependent on the assumed energy spectrum because they are
calculated differentially for narrow energy bins.  In order to compare
these values with other published (integral) upper limits, a differential power
law type spectrum has been assumed.  Notably, the HEGRA upper limits cover a
wider energy range than the other results and reach the lowest value in this
range. Furthermore, the upper limits are independent for each energy band
whereas all other quoted upper limits are integral limits which are less sensitive 
to narrow
features in the spectrum.

 The model predictions for outer gap emission as calculated recently by
 \citet{2001ApJ...558..216H} are partially excluded by the upper limits
 presented here. However, the expected flux from the Crab pulsar is strongly
 depending upon the opacity of the acceleration site for very high energy
 photons.  The opacity is governed by the ambient soft photon density and the
 strength of the magnetic field.  The optimistic cases considered with the gap
 sufficiently far away from the pulsar and low ambient photon density are
 nominally excluded by the upper limits given here.  As has been pointed out
 (Hirotani, private communication) further absorption of photons by single
 photon-pair production is expected and will reduce the expected flux
 considerably. The predictions of this model need to be revised and will not be
 constrained by this observation.

\section{Model calculations for the emission from the nebula and Discussion}
\label{section:model}
\subsection{Broad band spectral energy distribution}

 For the purpose of calculating the inverse Compton scattered radiation,  three basic 
photon fields need to be taken into account:
\begin{itemize} 
\item
Synchrotron emission: This radiation field dominates in density for all
energies and is the most important seed photon field present. 

 \item Far infrared excess: Observations at far-infrared wavelengths have shown the presence of possibly thermal emission
which exceeds the extrapolation of the continuum emission from the radio band. This component is best
described by a single temperature of 46~K \citep{1992Natur.358..654S}. Unfortunately, the spatial structure
of the dust emission remains unresolved which introduces uncertainties for the model calculations.
 We have assumed the dust to be distributed like the filaments with a scale-length of 1.3\arcmin. 
Sophisticated analyses of data taken with the ISO satellite indicates that the dust emission can be
resolved (Tuffs, private communication). The resulting size seems to be consistent with the value assumed here.

\item Cosmic Microwave Background (CMB):  Given the low energy of the CMB photons, scattering continues to take place
in the Thomson regime even for electron energies exceeding 100~TeV \citep{1995APh.....3..275A}.
\end{itemize}

The influence of stellar light has been found to be negligible \citep{1996MNRAS.278..525A}. 
The optical line emission of the filaments is spatially too far separated from the inner region of the nebula where
the very energetic electrons are injected and cooled. However, in the case of acceleration taking place at different
places of the nebula, the line emission could be important.

Given the recent detection of a compact component emitting mm radiation
\citep{2002A&A...386.1044B} this radiation field is included as seed photons for the calculation
of the inverse Compton scattering.
A simple model calculation has been performed which follows the phenomenological approach suggested by
\citet{1998ApJ...503..744H}. 

In brief, the observation of the continuum emission from the nebula up to MeV
energies is assumed to be synchrotron emission. By setting the magnetic field
to a constant average value within the nebula, a \textit{prompt} electron
spectrum can be constructed that reproduces the observed spectral energy
distribution. Based upon the measured size of the nebula at different
wavelengths, the density of electrons and the produced synchrotron photons can
be easily calculated in the approximation that the radial density profile follows a
Gaussian distribution. 

 With this simple model, it is straight-forward to introduce additional
electron components and seed photon fields to calculate the inverse Compton
scattered emission of the nebula.  The model is described in more detail by
\citet{horns}. 

\begin{figure}
\plotone{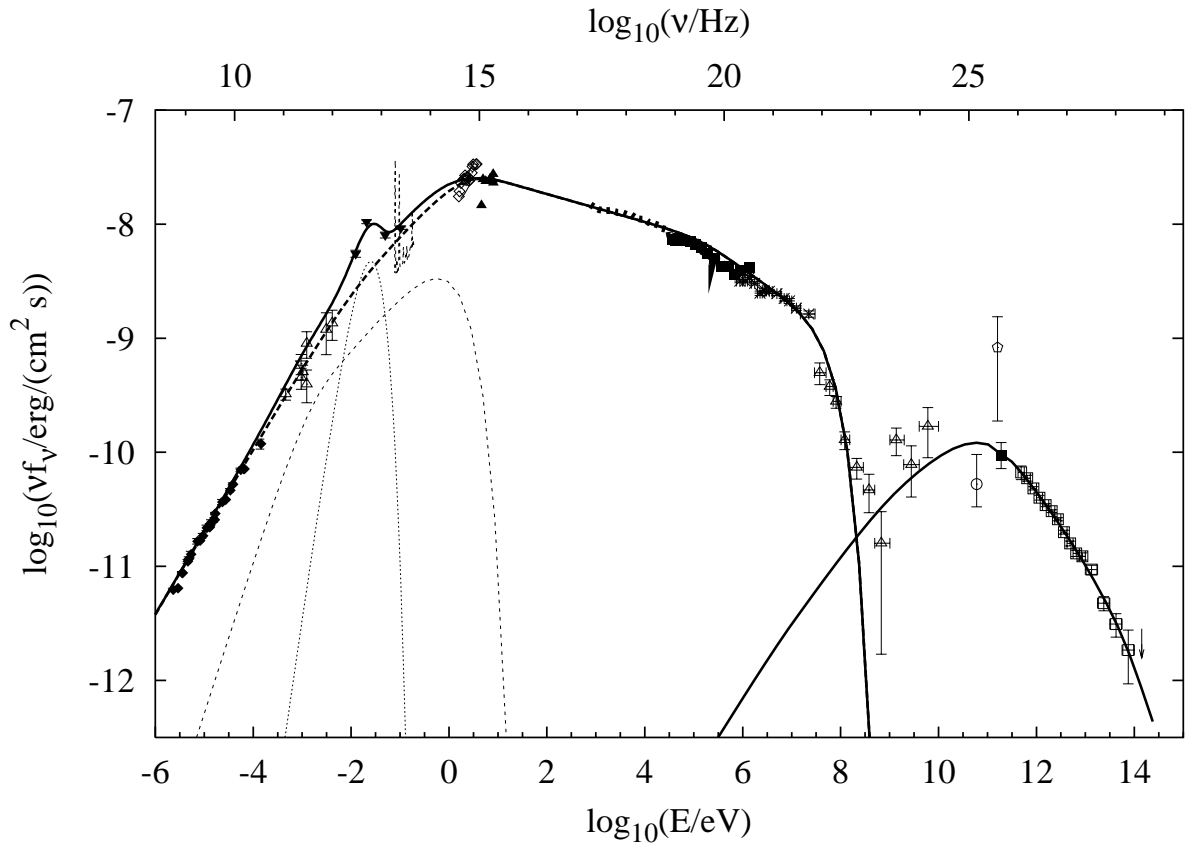}
 \caption{\label{fig:sed} For a wide range of energies, recent measurements have been 
compiled from the literature 
(see the text for further details and references). The curves are
calculations described in Section~\ref{section:model}. }
\end{figure}

In order to extract the underlying electron spectrum, a 
broad band SED is required (see Fig.~\ref{fig:sed}).
 For the purpose of compiling  and selecting available measurements
in the literature, mostly recent measurements have been chosen.  The prime goal
of the compilation is to cover all possible wavelengths from radio to
gamma-ray. The radio data are taken from \citep{1972A&A....17...172} and
references therein, mm data from
\citep{1986A&A...167..145M,2002A&A...386.1044B} and references therein, and the
infrared {data obtained with IRAS in the far- to mid-infrared  \citep{1992Natur.358..654S} and with ISO in the adjacent mid- to near-infrared band \citep{2001A&A...373..281D}.} 

 Optical and near-UV data from the Crab nebula require some extra
considerations. The reddening along the line-of-sight towards the Crab nebula
is a matter of some debate. For the sake of homogeneity, data in the optical
\citep{1993A&A...270..370V} and near-UV and UV \citep{1992ApJ...395L..13H,
1981ApJ...245..581W} have been corrected applying an average extinction curve
for $R=3.1$ and $E(B-V)=0.51$ \citep{1989ApJ...345..245C}.

 The high energy measurements of the Crab nebula have been summarized recently
in \citet{2001A&A...378..918K} to the extent to include ROSAT HRI, BeppoSAX
LECS, MECS, and PDS, COMPTEL, and EGRET measurements covering the range from
soft X-rays up to gamma-ray emission. For the intermediate range of hard X-rays
and soft gamma-rays, data from Earth occultation technique with the BATSE
instrument have been included \citep{2003ApJ...598..334L}. 

 The observations of the Crab nebula at very high energies (VHE, $E>100$~GeV)
have been carried out with a number of ground based detectors. Most successfully,
Cherenkov detectors have established the Crab nebula as a standard candle in
the VHE domain.
A summary of the measurements is
presented in \citet{2000ApJ...539..317A}. Recently, the MILAGRO group has published
a flux estimate which is consistent with the measurement presented here \citep{2003ApJ...595..803A}.

The results from different detectors reveal underlying
systematic uncertainties in the absolute calibration of the instruments.  To
extend the energy range covered in this work (0.5-80 TeV), results from
non-imaging Cherenkov detectors  like CELESTE {(open circle)},  STACEE {(solid square)}, 
and GRAAL {(open diamond)} at lower energy
thresholds have been included
\citep{2002ApJ...566..343D,2001ApJ...547..949O,2002APh....17..293A}, converted
into a differential flux assuming a power law for the differential energy
spectrum with a photon index of $2.4$. 
For energies beyond 100~TeV, an upper
limit on the integral flux  from the CASA-MIA air shower array has been added
\citep{1997ApJ...481..313B} assuming a power law with a photon index of $3.2$
as predicted from the model calculations.

 The resulting broad band SED is shown in Figure~\ref{fig:sed} including
as solid lines the synchrotron {and inverse Compton} emission as calculated with the electron
energy distribution assumed in this model.  Also indicated as a dotted line in Fig.~\ref{fig:sed} is the thermal excess radiation which is assumed to follow a modified 
black body radiation distribution with a temperature of 46~K. Finally, 
the emission at mm-wavelengths is indicated by a {thin} broken line 
(see also Section~\ref{section:mm}). {The thick broken line indicates the synchrotron
emission excluding the thermal infrared and non-thermal mm-radiation. The inverse
Compton emission shown in Fig.~\ref{fig:sed} includes the contribution
from mm emitting electrons (see next section).}

 Besides the spectral energy distribution, an estimate of the volume of the
emitting region is required to calculate the photon number density in the
nebula to include as seed photons for inverse Compton scattering. The size of
the nebula is monotonically decreasing with increasing photon energies which is
predicted in the  framework of the MHD flow models
\citep{1996MNRAS.278..525A,2000A&A...359.1107A}.

  The result of the calculation for the inverse Compton emission are
tabulated separately for the different contributions from the seed photon 
fields in Table~\ref{table:model}. {The table lists the differential flux
in the commonly used units photons\,$\mathrm{TeV}^{-1}\mathrm{cm}^{-2}\mathrm{s}^{-1}$ which
can be converted directly into energy flux by multiplying with the squared energy}. 
For convenience, Table~\ref{table:fit}
gives the coefficients for a polynomial parameterization of the energy
flux as a function of energy in a double logarithmic representation with
a relative linear accuracy better than 4~\%.

\subsection{Impact of a compact emission region at mm-wavelengths} \label{section:mm}
The recently found emission region at mm-wavelengths
\citep{2002A&A...386.1044B} is more compact than the radio and optical emission
region.  In the current calculation, this emission region is modeled by an
additional synchrotron emission component with a Gaussian equivalent radial
width of 36\arcsec\ as derived from Fig.~5 of \citep{2002A&A...386.1044B}.
The electrons radiating at mm-wavelengths produce also an additional inverse
Compton component at GeV energies which partially reduces the discrepancy
between the observed and predicted gamma-ray emission between 1 and 10 GeV.  As
a note, in order to match the EGRET flux, the mm-emission region would have to
be roughly half ($\approx 16$\arcsec) of the size used for this
calculation.

 As pointed out in \citet{2002A&A...386.1044B}, a thermal origin is ruled out
for several reasons which leaves the possibility of synchrotron emission. The
origin of the mm emitting electrons is unclear. Contrary to the common picture
of a relic electron population that was possibly emitted at an earlier stage of
the nebula's development (see also Section V of \citet{1984ApJ...283..710K} for
a discussion on the problems to produce such a radio component in the framework
of the MHD flow model), the mm electrons are apparently confined to or possibly
injected  into a similar volume as the soft X-ray emitting electrons. 

The extra component can be explained by a population of
electrons with a minimum Lorentz factor of $\approx 10^4$ and a maximum Lorentz
factor of $\approx 10^6$ with a power law index $p$ close to 2
($dN/d\gamma=N_0\,\gamma^{-p}$).  In order to inject such an electron
distribution, a small scale shock is required at 
a distance of $\approx 10^{14}$~cm
distance to the pulsar. The maximum Lorentz factor reachable in the downstream
region scales with the curvature radius of the shock. The minimum Lorentz
factor according to the Rankine-Hugoniot relations for a shock at $10^{14}$~cm
is $\approx 10^{4}$ provided that the spectral index is $\approx 2$ for the
particle distribution. In recent magneto-hydrodynamic calculations
\citep{2002MNRAS.336L..53B,2003MNRAS...344..L93} the observed jet-torus morphology of the Crab
nebula is reproduced by invoking a modulation of the flow speed with
$\sin\theta^2$ ($\theta$ is the polar angle with respect to the rotation axis
of the pulsar). This assumption is motivated by the solution for the wind flow
in the case of an oblique rotator assuming a split monopole magnetic field
configuration \citep{1999A&A...349.1017B}.  According to the  calculation of
\citet{2003MNRAS...344..L93}, a multi layered shock forms. In the proximity of
the polar region the shock would form close to the pulsar and could be
responsible for the injection of mm-emitting electrons. 

This shock region has remained undetected because of the angular proximity 
(10 mas) to the pulsar and the fact that the continuum emission is
predominantly produced in the mm and sub-mm wavelength band. The 
inner shock is in principle visible with high resolution
observations with interferometers at mm wavelengths.

Intriguingly, there  is observational evidence for ongoing injection of
electrons radiating at 5~GHz frequency which show moving wisp-like structures
similar to the optical wisps \citep{1992ApJ...393..206B,2001ApJ...560..254B}.
The injection of radio- and mm-emitting electrons into the nebula could e.g. take place at 
additional shocks much closer to the pulsar than the previously assumed
$14$\arcsec.

  The additional compact mm component is of importance for the inverse Compton
component: {The 
electron population with $\gamma=10^4\ldots10^5$ introduced here to explain the  excess at
mm-wavelengths produces via inverse Compton scattering gamma-rays between 1 and 10~GeV. The
contribution is rather small (10~\%) but could easily become comparable to the other components
if the mm-emitting region is smaller than assumed here. In this case, the EGRET data points 
would be better described by the model.}
{Moreover, the mm component contributes seed photons} for inverse Compton scattering
in the Thomson regime which contributes substantially at energies of a few TeV.

\begin{figure}
\plotone{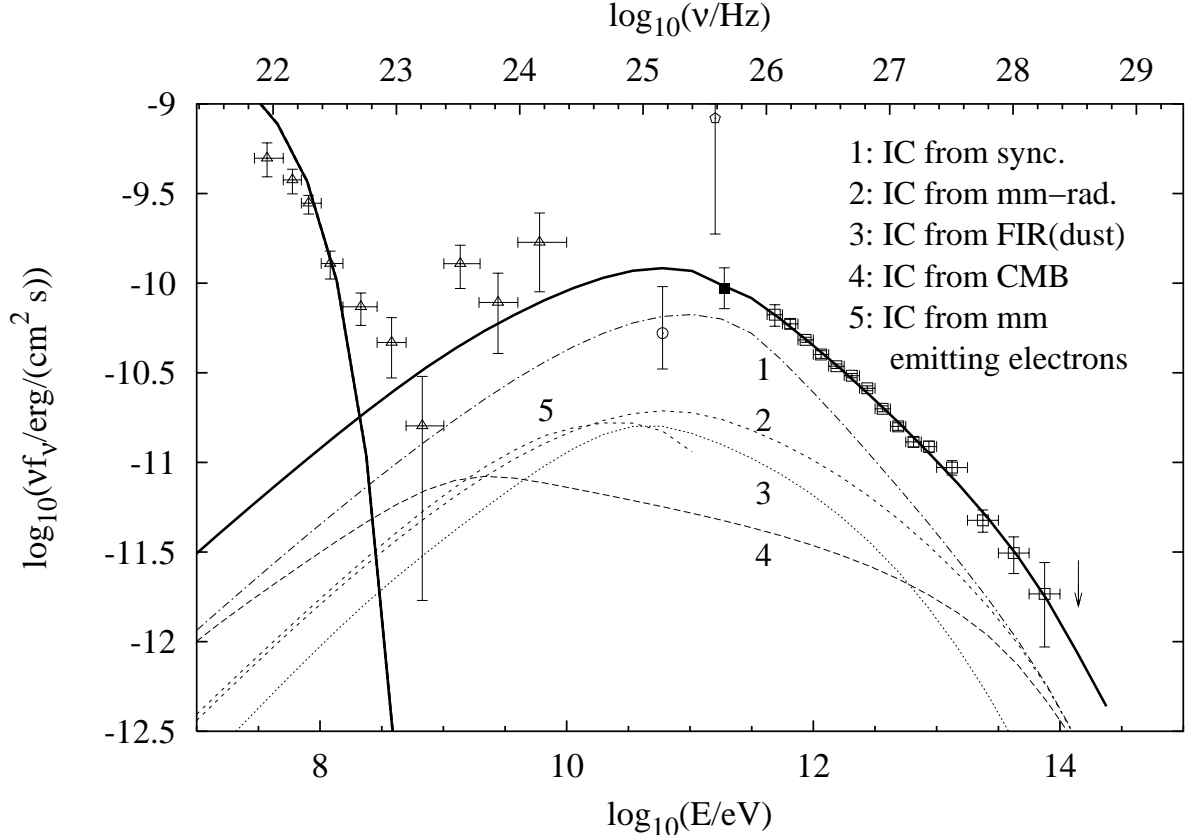}
\caption{\label{fig:ic} The inverse Compton spectrum is decomposed into different components as
defined by the target seed photons. The synchrotron radiation is dominating at all energies. However, the
mm-excess is of equal importance above 30~TeV. Eventually, beyond 100~TeV, the microwave background is
contributing significantly to the overall spectrum.  Note, the contribution of the mm-radiation is dominating 
over the dust component at all energies. Also shown is the inverse Compton component from the electrons
emitting the synchrotron mm-radiation. {The symbols used are the same as for Fig.~\ref{fig:sed}}
 }
\end{figure}

\begin{figure}
\plotone{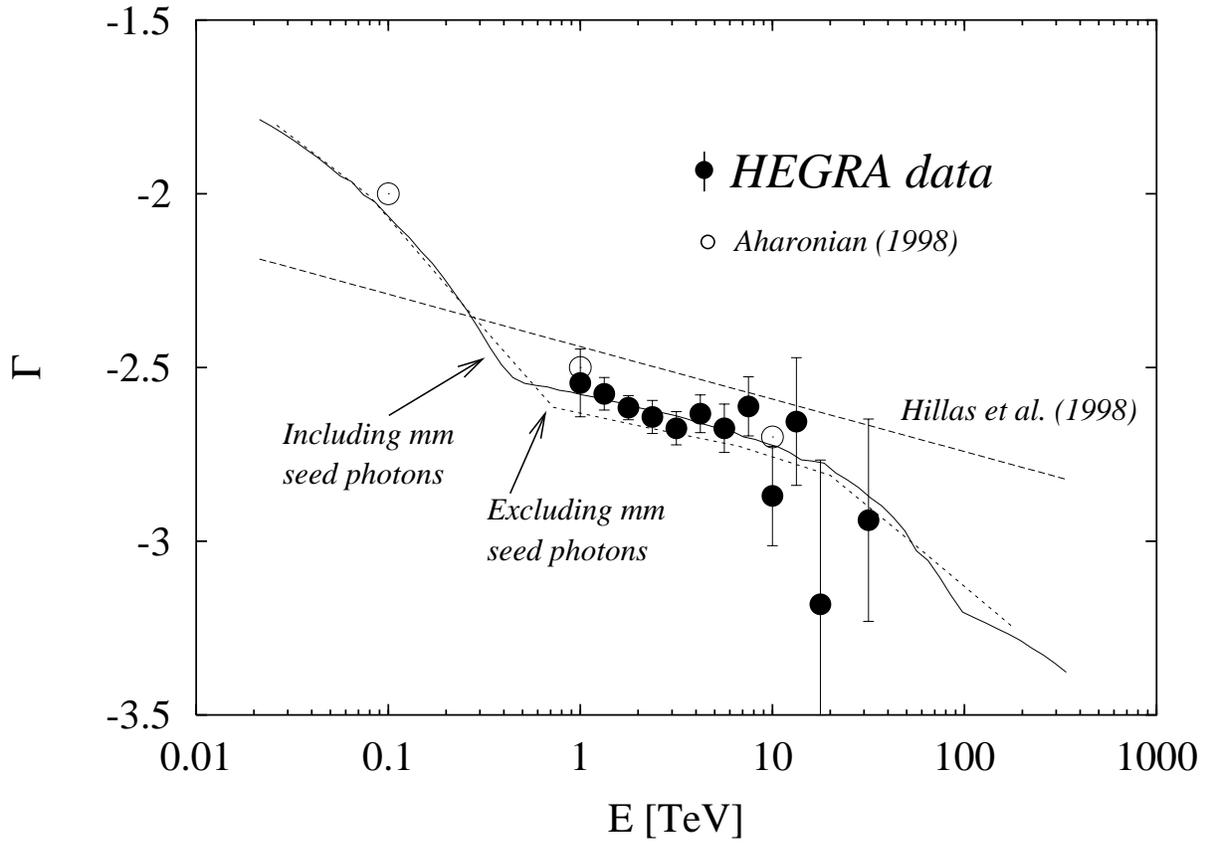}
\caption{\label{fig:diff} The predicted photon index (line) changes only very little in the 
energy range considered and is consistent with the measured values (filled symbols). 
}
\end{figure}

The combined inverse Compton spectrum is shown in Fig.\ref{fig:ic} decomposed
according to the different seed photons (Sy, IR, mm, CMB) {and the additional
inverse Compton component from the mm-emitting electrons}. Clearly, the
synchrotron emission present in the nebula is the most important seed photon
field present. At high energies ($E>10$~TeV) 
the mm-radiation is contributing significantly to the scattered radiation. 

\subsection{Comparison of the model with data}
\label{section:Comparison}
 The agreement of the calculated inverse Compton spectrum with the data
is excellent (see Fig.~\ref{fig:ic}). The only free parameter of the model calculation is the magnetic
field which in turn can be determined from the data 
by minimizing the $\chi^2$ of
the data (see Section~\ref{section:average_magnetic}). The resulting 
value of $\chi^2$ is slightly lower than for the power-law fit:
$\chi^2_{red}(d.o.f)=0.96(14)$ for the inverse Compton model
as compared with $\chi^2_{red}(d.o.f.)=1.3(13)$. However, the small 
change in $\chi^2$ does not convey the full information. The slope 
of the spectrum is expected to change slightly with energy for the 
inverse Compton model. For the purpose of testing this gradual softening 
predicted in the model, the differential power-law was calculated from the
data points by computing the slope between two data-points with index $i,j$ separated by 0.625
in decadic logarithm. 
\begin{eqnarray}
 \Gamma(E)     &=& \frac{\ln\Phi_i-\ln\Phi_j}{\ln E_i-\ln E_j}\\
 E	       &=& \exp(0.5\cdot(\ln E_i+\ln E_j))\\
 \sigma_\Gamma &=& \frac{\sqrt{(\sigma_{\Phi(i)}/\Phi_i)^2+(\sigma_{\Phi(j)}/\Phi_j)^2}}{\ln E_i-\ln E_j} 
\end{eqnarray}

For the sake of simplicity, the error on $E_i,E_j$ is ignored. The effect of including the statistical error on $E_i,E_j$ is negligible.
The interval of 0.625 in decadic logarithm gives sufficient leverage to 
calculate a reliable slope and at the same time it is small enough to resolve
the features. The expected slope is calculated from the model for both cases
including
and excluding the mm seed photon field. The result is shown in 
Fig.~\ref{fig:diff}. The straight broken line is the predicted slope as 
given by the parameterization of \citet{1998ApJ...503..744H}, the open symbols indicate
the prediction from \citet{aharonian...nsp} which are consistent with the calculation described 
here.
 Note, the data points are independent. As is clearly
seen, the expected and measured change in slope agree nicely. The systematic 
and energy dependent deviation from the constant photon index determined 
by the power law fit is evident. Ignoring the mm-component gives on 
average a slightly softer spectrum with the same softening with increasing 
energy. It is
remarkable how little the slope is expected to change over exactly the
energy range covered by the observations. Going to energies below 200~GeV, 
a strong flattening of the energy spectrum is to be expected. At the
high energy end, beyond roughly 70~TeV the softening is expected to 
be stronger.

\subsection{The average magnetic field}
\label{section:average_magnetic}
 The average magnetic field is calculated by minimizing the $\chi^2$ 
of the model with respect to the data varying the magnetic field as a free parameter.
For every value for the magnetic field, the break energies and normalization
of the electron spectrum is chosen to reproduce the synchrotron spectrum.

The resulting $\chi^2$ as a function of magnetic field shows a minimum at 
$161.6~\mu$G with $\chi^2_{red}(d.o.f.)=0.96(14)$. The  
 $1\sigma$ statistical uncertainty of $0.8~\mu$G is negligible. The systematic
uncertainties of $15~$\% on the absolute energy scale of the measurement translates into 
a systematic error on the average magnetic field  of 18~$\mu$G.

 %The normalization of the 
%inverse Compton component is proportional to the inverse Compton luminosity $L_{IC}$.
%Under the assumption that the observed emission at lower energies is synchrotron 
%emission of the same electrons with a fixed luminosity $L_{Sy}\propto nB^2$ with 
%$n$ being the electron number density and $L_{IC}\propto n$ (keeping the 
%seed photon density fixed), the average magnetic field  $B\propto \sqrt{L_{Sy}/L_{IC}}=
%\sqrt{f_{Sy}/f_{IC}}$ independent of the distance to the object.  For the sake of 
%getting an estimate on the 
% Given the very small statistical error
%on the normalization of 1.1~\%, the magnetic field is in principle determined with
%a statistical uncertainty of 0.55~\%. However, the systematic uncertainty on the energy
%scale introduces a systematic error on the magnetic field of 12~\%. 
%The minimum $\chi^2=14.7$ with 16 degrees of freedom is showing the excellent agreement
%of the model and the data. The magnetic field is determined to be $(159\pm0.8_\mathrm{stat}\pm 17_\mathrm{sys})~\mu$G. 
%

\subsection{Gamma-ray emission from ions in the wind}
\label{section:ions}

 As has been pointed out independently by \citet{2003A&A...402..827A} and
\citet{2003A&A...405..689B} the presence of ions in the relativistic wind could
be detected by the production of neutrinos and gammas in
inelastic scattering processes with the matter in the nebula.  Neutrinos and gammas
would be produced as the decay products of charged and neutral pions.
Both calculations show a similar signature for the 
ion induced gamma-ray flux which appears as a rather narrow feature in a $\nu f_\nu$ 
diagram. 
The presence of ions is required in acceleration models in which positrons in the downstream region are
accelerated by cyclotron waves excited by ions \citep{1995mpds.conf..257A}.

Qualitatively, the ions in the pulsar wind fill the nebula without 
undergoing strong energy losses.  For a typical bulk velocity of the wind of 
$\Gamma=10^6$, the dominant energy losses are adiabatic expansion
of the nebula and escape of particles with typical time scales of the order
of the age of the remnant. Therefore, the almost monoenergetic distribution of
the ion energies as it is injected by the wind is not widened significantly in energy.

 From the model calculations described above, the TeV data are well explained by
inverse Compton scattering of electrons present in the nebula. The shape and
the absolute flux measured is consistent with the prediction of this model.
However, an admixture of gamma-ray emission processes of electrons and ions  could
be possible. In an attempt to estimate how much of the spin down power of the
pulsar could be present in the form of ions, a model calculation for the
gamma-ray emission from ions in the wind is performed. Here, 99~\% c.l.
upper limits are calculated on the fraction $\zeta$ of the spin down luminosity present in
ions ($L_{ion}=\zeta \dot E_{pulsar}$) by adding the additional component to
the inverse Component component and comparing this prediction with the data.
The $\chi^2$ of the combined prediction is calculated for different values of
$\zeta$ until the $\chi^2$ increases by $\Delta \chi^2=29.15$ for a 99~\%
confidence level.  This calculation is repeated for various values of $\Gamma$.

 The calculation of the gamma-ray flux from the ions depends upon a few
simplifications.  It is assumed that the ions are predominantly protons 
with a narrow energy distribution with $E_p\approx\Gamma m_pc^2$.  For the injection rate of protons $\dot
N_p=\zeta \dot E_{pulsar}/E_p$ with $\dot E_{pulsar}=5\cdot 10^{38}$~erg\,s$^{-1}$ is
assumed.  The injection rate of $\pi^0$ is depending on the conversion
efficiency: $\dot N_{\pi}=f_\pi \dot N_p$. The conversion efficiency $f_\pi$ of
the protons to $\pi^0$ is given by the ratio of energy loss time scales for
pp-scattering and escape/adiabatic losses. Assuming an average number density of
the gas in the nebula of 5 cm$^{-3}$, the conversion efficiency is
$5.4\times10^{-5}$. 
% The $\pi^{0}$ decay spectrum is calculated with a scaling
%approximation (see eg. \citet{2001APh....15..223B}).  

\begin{figure}
\plotone{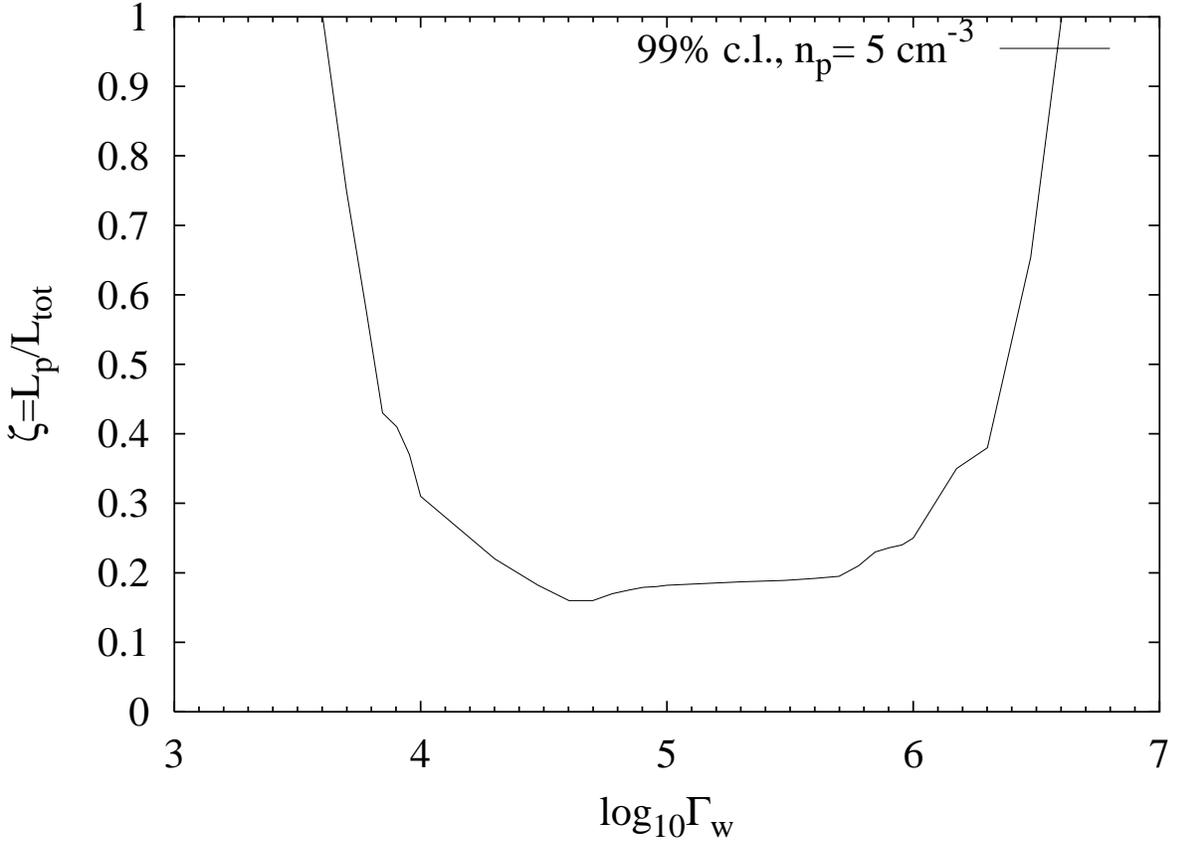}
\caption{\label{fig:zeta} For a variety of bulk Lorentz speeds of the wind, 
the fraction of the total spin down power which is injected in the form of
ions is constrained to be less than 20~\%. For Lorentz factors beyond $10^{6}$ and
below $10^4$ a substantial fraction of the power could be present in the form 
of ions.}
\end{figure}

  The upper limits for $\zeta$ as a function of the bulk Lorentz factor are
given in Fig.~\ref{fig:zeta}. Clearly, for bulk Lorentz factors between $10^4$
to $10^6$, only a small fraction of the energy carried by the wind could be
present in the form of ions ($<20$~\%). For Lorentz factors beyond $10^6$ or
smaller $10^4$ a substantial part of the power could be injected in the
form of ions.

   However, the assumption of a narrow energy distribution might not be true.
In the case of a variation of the wind speed as discussed above in the context
of a split magnetic monopole model, the injected ions could have a wider
distribution in energy which subsequently makes an identification of this
feature in the measured gamma-ray spectrum more complicated and upper limits
calculated less restrictive.

\subsection{Additional Components}
\label{section:more}

 The presence of additional electron populations in the nebula with $\gamma>10^5$ can 
be constrained by the gamma-ray data presented here. Obviously, the observed synchrotron 
emission already excludes the presence of components not accounted in the model calculations
presented in Section~\ref{section:model}.  However, the frequency interval between hard UV and soft X-ray emission
is not covered by observations and large systematic uncertainties on the strong absorption present would
in principle not exclude the presence of an additional component that is sufficiently narrow in energy.  Specifically,
electrons following a relativistic Maxwellian 
type distribution may form in the framework of specific acceleration models \citep{1992ApJ...390..454H}. 
For the magnetic field of 161~$\mu$G, UV-emitting electrons would have $\gamma\approx 10^7$. Given the age 
of the nebula of 954 years, the synchrotron-cooled spectrum is close to a power-law with $dN/d\gamma\propto \gamma^{-2}$ reaching
to $\gamma_{min}=10^6$. The resulting synchrotron emission is not in conflict with the available data unless 
the power in the extra Maxwellian type injection spectrum  becomes a sizable fraction ($L_{Maxw.}/L_{tot}>0.3$) 
of the total power of the
electrons responsible for the broad-band SED. 
However, after calculating the inverse Compton emission of this extra component and comparing it with the
gamma-ray data,  an upper limit at 99~\% c.l. can be set $L_{Maxw.}/L_{tot}<0.04$ ruling out the existence of unnoticed
(unobserved) electron populations with $\gamma=10^6\ldots10^7$ in the nebula
The possible existence of an additional 
 synchrotron emitting component possibly associated with a spectral hardening in the MeV energy 
range as observed consistently by the COMPTEL instrument \citep{1995A&A...299..435M},
BATSE \citep{2003ApJ...598..334L}, and INTEGRAL \citep{2003A&A...411L..91R}
and has been discussed e.g. in \citet{gamma2000_jager}. The contribution of such an additional component to the 
observed TeV flux is probably small and is relevant only at energies exceeding a few 100 TeV.

\section{Summary and Conclusions}
\label{section:summary}

 The HEGRA stereoscopic system of air Cherenkov telescopes performed extensive
observation of the Crab supernova remnant. The energy spectrum has been
determined over more than 2 decades in energy and 5 decades in flux. 
 The main points of the data analyses are to constrain the acceleration
of electron/positron pair plasma in the vicinity of the termination shock
of the pulsar wind in the surrounding medium. The high energy data 
presented here give a unique view on the extreme accelerator which resides
in the Crab nebula. 

 The energy spectrum of the unpulsed component is very likely produced by the
same electrons that produce the broad-band emission spectrum extending from
soft to hard X-rays and finally reach the soft gamma-rays (with a cut-off at  a
few MeV). The electrons up-scatter mainly photons of the synchrotron nebula,
the soft thermal photons seen as the far infrared excess, the universal cosmic
microwave background (CMB), and possibly mm-radiation emitted in a rather
compact region.

 The comparison of the flux level at TeV energies combined with
the synchrotron flux in a simple spherical model, constrains directly the
magnetic field in the emission region of the 
nebula to be at the level of 160~$\mu$G.  The agreement
of the measured broad band energy spectrum ranging from soft X-rays up to 100~TeV
with the prediction based upon  a simple model of an electron population being
injected at the standing reverse shock strengthens the claim that electrons
with energies exceeding $10^{15}$~eV are continuously accelerated.  Limited by
the photon statistics detected at high energies ($>10$~TeV), expected variations
in the acceleration/cooling rate on time scales of months for these electrons
can not be established with the current instrumentation. 

 The {data show evidence for the} predicted gradual softening of the energy spectrum.
 Below roughly 200~GeV the spectrum is expected to harden quickly. 
Future detectors with a sufficiently low energy threshold will be able to
see this feature. However, the expected \textit{softening} occurs rapidly at energies
beyond 70~TeV. Here, low elevation observations with the Cherenkov
telescopes from the southern hemisphere will be very helpful.
% The collection
%area for a H.E.S.S. type system of Cherenkov telescopes reaches values 
%close to a 1~km$^2$ at these energies at elevation angles of 20$^\circ$.

 Given the good agreement of the predicted inverse Compton spectrum with the
 measurement, the presence of ions in the wind is either negligible or
 radiative losses are exceedingly small. 
  Independent of the spectrum, the morphology
 of the emission region could reveal signatures of the presence of ions in the
 wind and the nebula. Experimental upper limits given here constrain the size
 of the emission region at the considered energies to be less than 2\arcmin\ and
 less than 3\arcmin\ above 10~TeV.  This excludes for example a strong
 contribution from the outer shock of the (undetected) expanding super nova
 remnant and constrains the transport of ions in the wind.

 Finally, pulsed emission from inside the pulsar's magnetosphere at high
 energies is expected in outer gap models.  A dedicated search for narrow
 features as predicted in this type of model has been performed and in the
 absence of a signal, upper limits have been calculated. The upper limits reach
 well below the more optimistic predictions of \citet{2001ApJ...558..216H} but more
 recent calculations indicate that the pair opacity in the emission region is
 possibly larger than originally anticipated.  In a different scenario, pulsed
 emission is expected to arise as a consequence of inverse Compton scattering
 of pulsed soft photons by the un-shocked wind \citep{2000MNRAS.313..504B}. The
 upper limits derived here are useful in constraining combinations of the bulk
 Lorentz factor and the distance of wind injection to the pulsar. For a range
 of bulk Lorentz factors from $10^5-10^7$ the distance of the wind is
 constrained to be accelerated more than 50 light radii away from the pulsar.

\section{Outlook}
 Future observations of the nebula with low energy threshold instruments (MAGIC, GLAST) and large 
collection area observations at small elevations from the southern hemisphere (CANGAROO III, H.E.S.S.)
should extend the accessible energy range below 100~GeV  (MAGIC) and beyond 100~TeV (CANGAROO III, H.E.S.S.) 
where the spectral shape is expected to deviate strongly from a power-law with 
$\Gamma=-2.6$. With low energy threshold Cherenkov telescopes like MAGIC the detection of pulsed emission 
from the pulsar becomes feasible. Combining information about possible variability of the nebula's emission at different
wavelengths (specifically hard X-rays and say 20-100 TeV emission) would ultimately prove the common (electronic) origin of
the observed emission.    Fortunately, the Crab will be frequently observed as a calibration source by all 
Cherenkov telescopes  and by INTEGRAL and other future hard X-ray missions so that even without dedicated observations, 
substantial observation time will be accumulated.  
A dedicated mm-observation with sub-arcsecond resolution of the region of the nebula within
the X-ray torus would be of great interest in order to confirm the predicted existence of multi-layered shocks which could
be responsible for the mm-emission detected with the moderate ($>$ 10\arcsec ) resolution
observations currently available. If the injection
of mm-emitting electrons in a compact region is confirmed, inverse Compton scattered emission produced by the same
low energy electrons could in principle explain the {discrepancy of the model with the
energy spectrum} observed by the EGRET spark-chamber onboard the
CGRO satellite. Again, future observations with the GLAST satellite with improved statistics will be of great interest to
study the actual shape of the energy spectrum above a few GeV. 
\begin{acknowledgements}
The support of the German ministry for research and
technology (BMBF) and of the Spanish Research Council (CICYT) is gratefully
acknowledged. GR acknowledges receipt of a Humboldt fellowship.
We thank the Instituto de Astrof\'{\i}sica de Canarias
for the use of the site and for supplying excellent working conditions at
La Palma.  We thank K. Hirotani, Y.~Gallant, S. Bogovalov, D. Khangoulian, 
and O. Skj{\ae}raasen for inspiring discussions. We acknowledge the
support of the Jodrell Bank pulsar team for providing us with the
ephemerides of the Crab pulsar and A. Franzen for letting us
analyze the optical data taken with the H.E.S.S. telescopes on the Crab pulsar.
This research has made use of NASA's Astrophysics Data System Bibliographic Services and
the Chandra X-Ray Observatory Science Center. {We thank the anonymous referee for his
suggestions to improve the manuscript.}
\end{acknowledgements}

\clearpage
%%%%%%%%%%%%%%%%%%%%%%%%%%%%%%%%%%%%%%%%%%%%%%%%%%%%%5
\begin{deluxetable}{lccccc}
\tablewidth{0pt}
\tablecaption{Observational times
 \label{table:data_selection}
 }
% (after selection on good weather) for different zenith angle 
%bins}
\tablehead{
\colhead{Season} & 
\colhead{$[6.5^\circ,30^\circ[$} & 
\colhead{$[30^\circ,45^\circ[$} & 
\colhead{$[45^\circ,65^\circ]$} &
\colhead{$\sum$} & 
\colhead{$\sum$} \\
year   &  $[$ksec$]$                     & $[$ksec$]$                  & $[$ksec$]$                  
        & $[$ksec$]$ & $[$hrs$]$ 
}
\startdata
1997/98 &165.68 &125.47 &75.94  &367.08 &  101.97\\
1998/99 &217.89 &91.54	&177.23	&486.66 &  135.18\\
1999/00 &89.35	&28.67 	&0.81	&118.83	&   33.00\\
2000/01	&113.21	&67.46	&6.40	&187.08	&   51.97\\
2001/02	&95.94	&100.30	&29.61	&225.84	&   62.73\\
\hline
Total   &682.07	&413.43	&289.99	&1385.49&  384.86
\enddata
\tablecomments{Data after selection for good weather}
\end{deluxetable}

\begin{deluxetable}{lccc}
 \tablewidth{0pt}
 \tablecaption{Image and event selection for different analyses
\label{table:selection}}
 \tablehead{
 \colhead{Selection} &
 \colhead{Spectrum} &
 \colhead{Morphology} &
 \colhead{Pulsation}}
% The values for the core and 
%stereo angle cut in parentheses refer to large zenith angle observations (z.a.$>45^\circ$).
%For the studies on the source morphology, two different cuts on the estimated uncertainty
%of the reconstructed direction are applied: $\sigma_{xy}<0.1^\circ$ is used for finding the
%position of the emission region and $\sigma_{xy}<0.05^\circ$ is used for constraining the 
%size of the emission region.
\startdata
Angular distance to source & $\vartheta<0.225^\circ$       &  no cut                   & $\vartheta<0.118^\circ$     \\
Image shape  & $\tilde w<1.1$      &$\tilde w<1.1$      & $\tilde w<1.1$         \\
Image amplitude		& $size>40$~p.e.      &$size>40$~p.e.      & $size>40$~p.e.          \\
Image centroid position 	        & $distance<1.7^\circ$& no cut & no cut \\
\hline
Estimated error on rec.                & no cut              &$\sigma_{sx}<0.1^\circ(0.06^\circ)$ \tablenotemark{a} &  no cut \\ 
event direction                        &                     &                                                      & \\
Core position  & $r<200(400)\tablenotemark{b}$~m		       & no cut  & no cut                     \\
Stereo Angle  & $min(stereo)>20^\circ(10^\circ)\tablenotemark{c}$  & no cut & no cut                     \\
Number of images      & $n_{image}\ge 3$		& $n_{image}\ge2$  & $n_{image}\ge2$  \\
Zenith angle      & $z.a.<65^\circ$                & $z.a.<65^\circ$   & $z.a.<65^\circ$ 
\enddata
\tablenotetext{a}{The cut on the estimated error for individual event direction $\sigma_{xy}<0.1^\circ$ is used when  
calculating the position of the emission region. The tighter cut ($\sigma_{xy}<0.06^\circ$) 
is applied to data used to constrain the size of 
the emission region}
\tablenotetext{b}{For large zenith angles, the cut on the impact distance is relaxed ($r<400$~m).}
\tablenotetext{c}{For large zenith angles, the cut on the minimum stereo angle subtended between the
major axes of the images in two telescopes is relaxed to $10^\circ$.}

\end{deluxetable}

\begin{deluxetable}{rrcrrr}
\tablecaption{Differential energy spectrum
\label{table:spec}
}
% statistical, and systematic errors after
%combining different zenith angle data. $N_{off}$ is the sum of 
%events counted in 5 background regions: $\langle N_{off}\rangle=0.2\,N_{off}$}
\tablewidth{0pt}
\tablehead{
\colhead{$\langle E\rangle$}  & 
\colhead{$E_\mathrm{low}-E_\mathrm{high}$} &
\colhead{$d\Phi/dE\pm \sigma_{stat}$} & 
\colhead{$N_{on}$}& 
\colhead{$N_{off}$\tablenotemark{a}}& 
\colhead{$S$ \tablenotemark{b}} \\
\colhead{$[\mathrm{TeV}]$} & 
\colhead{$[\mathrm{TeV}]$} & 
\colhead{$[(\mathrm{cm}^2\,\mathrm{s}\,\mathrm{TeV})^{-1}]$} &
 &
 & 
\colhead{$[\sigma]$}}
\startdata
       0.365 & 	0.316-0.422 & $(1.97\pm1.17)\cdot10^{-10}$ & 105  &333  & 3.9 \\
       0.487 &   0.422-0.562 & $(1.76\pm0.24)\cdot10^{-10}$ & 369  &705  & 14.1\\
       0.649 &   0.562-0.750 & $(8.78\pm0.53)\cdot10^{-11}$ & 1012 &1356 & 29.8 \\
       0.866 &   0.750-1.000 & $(4.02\pm0.13)\cdot10^{-11}$ & 2119 &2108 & 50.0 \\
       1.155 &   1.000-1.334 & $(1.87\pm0.09)\cdot10^{-11}$ & 2829 &2772 & 58.2\\
       1.540 &   1.334-1.778 & $(9.05\pm0.26)\cdot10^{-12}$ & 2458 &2220 & 56.1\\
       2.054 &   1.778-2.371 & $(4.51\pm0.12)\cdot10^{-12}$ & 2017 &1600 & 48.9\\
       2.738 &   2.371-3.162 & $(2.16\pm0.07)\cdot10^{-12}$ & 1510 &1114 & 47.3\\
       3.652 &   3.162-4.217 & $(9.33\pm0.36)\cdot10^{-13}$ & 950  &645  & 38.6\\
       4.870 &     4.217-5.623 & $(4.18\pm0.20)\cdot10^{-13}$ & 579  &330  & 31.7\\
       6.494 &   5.623-7.499 & $(1.93\pm0.12)\cdot10^{-13}$ & 345  &187  & 23.3\\
       8.660 &   7.499-10.000& $(1.02\pm0.07)\cdot10^{-13}$ & 238  &111  & 21.4\\
    13.335   &	10.000-17.783& $(3.28\pm0.31)\cdot10^{-14}$& 414  &420  & 19.7\\
    23.714   &     17.783-31.622&$(5.28\pm0.70)\cdot10^{-15}$ &  150 &242  & 10.2\\
    42.170   &	31.622-56.234&$(1.10\pm0.25)\cdot10^{-16}$ &  69  & 141 & 5.7\\
    74.989   &    56.234-100.000   &$(2.05\pm1.01)\cdot10^{-16}$ &36&104  & 2.7
\enddata
\tablenotetext{a}{The value $N_{off}$ is the sum of the background events in five discrete
background regions: $\langle N_{off} \rangle = N_{off}/5$}
\tablenotetext{b}{The significance is calculated for the number of excess events
$N_{on}-\langle N_{off}\rangle$ and invoking equation 17 of Li\&Ma(1983) with $\alpha=0.2$.}
\end{deluxetable}

\begin{deluxetable}{rrrr}
\tablewidth{0pt}
\tablecaption{Differential upper limits on pulsed emission
\label{table:pulsed_limits}
}

% with a 3$\,\sigma$ confidence level (99.865~\,\% c.l.) have
%been derived for individual energy bins.}
\tablehead{
\colhead{$\langle E\rangle$} &
\colhead{$E_{low}-E_{high}$} &
\colhead{fractional limit} &
\colhead{flux limit} \\
\colhead{$[\mathrm{TeV}]$} &
\colhead{$[\mathrm{TeV}]$} &
\colhead{$[\%]$} &
\colhead{$[10^{-14}$~ph\,(cm$^2$~s~TeV)$^{-1}]$}
}
\startdata
0.548 & 0.32 -  1.00   & 3.3 	& 392.5	\\
1.414 &  1.00 -  2.00   & 2.4	& 29.6	\\
2.828 & 2.00 -  4.00   & 2.8	& 6.03	\\
5.657 & 4.00 -  8.00   & 4.3	& 1.32	\\
11.314&  8.00 - 16.00   & 7.4	& 0.41	\\
22.627&  16.00 -  32.00 & 7.1	& 0.05	\\
56.569&  32.00 - 100.0  &43.4 	& 0.02	
\enddata
\end{deluxetable}

\begin{deluxetable}{rrrrrrr}
\tabletypesize{\footnotesize}
\tablewidth{0pt}
\tablecaption{Predicted
 inverse Compton differential energy spectrum for different seed photons. The
numbers (\textit{1-5}) refer to the component with the respective label in Fig.~10
\label{table:model} }
\tablehead{
\colhead{Energy} &
\colhead{Total} &
\colhead{\textit{1}} &
\colhead{\textit{2}}&
\colhead{\textit{3}} &
\colhead{\textit{4}} &
\colhead{\textit{5}}\\
\colhead{$\log_{10}(\mathrm{E}/\mathrm{TeV})$} &
\multicolumn{6}{c}{$\log_{10} (d\Phi/dE \cdot (\mathrm{cm}^2~\mathrm{s}~\mathrm{TeV}))$} 
}
\startdata
-2.3835 & -5.5682 & -5.9685  & -6.4142 & -6.5613 & -6.5423  & -6.1671 \\ 
-2.2168 & -5.8493 & -6.2340  & -6.6881 & -6.8169 & -6.8950  & -6.4603 \\ 
-2.0501 & -6.1358 & -6.5045  & -6.9685 & -7.0784 & -7.2499  & -6.7664 \\ 
-1.8835 & -6.4293 & -6.7811  & -7.2561 & -7.3481 & -7.6054  & -7.0891 \\ 
-1.7168 & -6.7305 & -7.0644  & -7.5515 & -7.6295 & -7.9614  & -7.4289 \\ 
-1.5501 & -7.0412 & -7.3557  & -7.8558 & -7.9268 & -8.3177  & -7.7880 \\ 
-1.3835 & -7.3618 & -7.6561  & -8.1704 & -8.2436 & -8.6747  & -8.1660 \\ 
-1.2168 & -7.6932 & -7.9668  & -8.4962 & -8.5827 & -9.0324  & -8.5616 \\ 
-1.0501 & -8.0372 & -8.2903  & -8.8345 & -8.9416 & -9.3908  & -8.9833 \\ 
-0.8835 & -8.3966 & -8.6298  & -9.1860 & -9.3135 & -9.7503  & -9.4553 \\ 
-0.7168 & -8.7994 & -8.9893  & -9.5500 & -9.6927 & -10.1109  &    \\ 
-0.5501 & -9.1819 & -9.3752  & -9.9242 & -10.0780 & -10.4731  &    \\ 
-0.3835 & -9.5879 & -9.7950  & -10.3056 & -10.4700 & -10.8368  &    \\ 
-0.2168 & -10.0108 & -10.2403  & -10.6928 & -10.8694 & -11.2027  &    \\ 
-0.0501 & -10.4374 & -10.6914  & -11.0862 & -11.2773 & -11.5710  &    \\ 
0.1165 & -10.8674 & -11.1465  & -11.4859 & -11.6941 & -11.9421  &    \\ 
0.2832 & -11.3010 & -11.6052  & -11.8925 & -12.1209 & -12.3166  &    \\ 
0.4499 & -11.7378 & -12.0673  & -12.3060 & -12.5583 & -12.6950  &    \\ 
0.6165 & -12.1785 & -12.5325  & -12.7267 & -13.0072 & -13.0782  &    \\ 
0.7832 & -12.6235 & -13.0008  & -13.1547 & -13.4685 & -13.4669  &    \\ 
0.9499 & -13.0730 & -13.4722  & -13.5905 & -13.9431 & -13.8621  &    \\ 
1.1165 & -13.5277 & -13.9470  & -14.0343 & -14.4318 & -14.2652  &    \\ 
1.2832 & -13.9885 & -14.4258  & -14.4868 & -14.9353 & -14.6784  &    \\ 
1.4499 & -14.4571 & -14.9100  & -14.9494 & -15.4541 & -15.1047  &    \\ 
1.6165 & -14.9366 & -15.4021  & -15.4246 & -15.9887 & -15.5496  &    \\ 
1.7832 & -15.4319 & -15.9060  & -15.9118 & -16.5408 & -16.0182  &    \\ 
1.9499 & -15.9477 & -16.4287  & -16.4345 & -17.1204 & -16.5060  &    \\ 
2.1165 & -16.4827 & -16.9703  & -16.9751 & -17.7492 & -17.0073  &    \\ 
2.2832 & -17.0254 & -17.5166  & -17.5249 & -18.3934 & -17.5230  &    \\ 
\enddata
\end{deluxetable}

\begin{deluxetable}{cccccc}
\tablewidth{0pt}
\tablecaption{Polynomial coefficients for a fit to the total inverse Compton 
energy flux (corresponds to the second column in Table~\ref{table:model}) of the form 
$\log_{10} \{\nu f_\nu /(\mathrm{erg}\,(\mathrm{cm}^2\,\mathrm{s})^{-1})\}
=\sum_{i=0}^5 p_i \log_{10}^i(E/\mathrm{TeV})$. \label{table:fit}}
\tablehead{
\colhead{$p_0$} &
\colhead{$p_1$} &
\colhead{$p_2$} &
\colhead{$p_3$} &
\colhead{$p_4$} &
\colhead{$p_5$}}
\startdata
-10.3531  &
-0.578559  &
-0.119778 &
0.542375$\cdot 10^{-1}$ &
-0.766819$\cdot 10^{-2}$  &
-0.660223$\cdot 10^{-2}$ 
\enddata
\end{deluxetable}

\end{document}